\documentclass[12pt]{iopart}

\pdfoutput=1

\usepackage{blindtext}

\usepackage[bottom]{footmisc}
\usepackage{float}
\usepackage{multirow}
\usepackage{array}
\newcolumntype{P}[1]{>{\centering\arraybackslash}p{#1}}
\usepackage{longtable}

\restylefloat{table}
\usepackage{lscape}
\usepackage{tabto}  
\usepackage[euler]{textgreek}
\usepackage{tabularx} 
 \usepackage{ragged2e}

\usepackage[hyphens]{url}
\usepackage{hyperref}
\usepackage[hyphenbreaks]{breakurl}

\begin{document}

\title{Anomalies in Radioactive Decay Rates: \\ A Bibliography of Measurements and Theory}

\author{M. H. McDuffie$^{1}$,  P. Graham$^{2}$, J. L. Eppele$^{3}$, J. T. Gruenwald$^{2}$, D.~Javorsek II$^{4}$, D.~E.~Krause$^{5,1}$, E.~Fischbach$^{1,2}$\footnote{Corresponding author (\mailto{ephraim@purdue.edu})}}

\address{$^{1}$ Department of Physics and Astronomy, Purdue University, West Lafayette, IN 47907 USA}
\address{$^{2}$ Snare, Inc., West Lafayette, IN 47906, USA}
\address{$^{3}$ School of Electrical and Computer Engineering, Purdue University, West Lafayette, IN 47907 USA}
\address{$^{4}$ U.S. Air Force, Arlington, VA 22203}
\address{$^{5}$ Department of Physics, Wabash College, Crawfordsville, IN, 47933 USA}

\vspace{10pt}
\begin{indented}
\item[]\today
\end{indented}

\begin{abstract}
Knowledge of the decay rates (or half-lives) of radioisotopes is critical in many fields, including medicine, archeology, and nuclear physics, to name just a few. Central to the many uses of radioisotopes is the belief that decay rates are fundamental constants of nature, just as the masses of the radioisotopes themselves are. Recently, the belief that decay rates are fundamental constants has been called into question following the observation of various reported anomalies in decay rates, such as apparent periodic variations. The purpose of this bibliography is to collect in one place the relevant literature on both sides of this issue in the expectation that doing so will deepen our understanding of the available data. 
\end{abstract}

\section{Introduction}

In recent years numerous experiments have presented evidence questioning whether  decay rates of radioactive nuclei, or equivalently their half-lives, are fundamental constants of nature, as is generally believed. As examples, data from an experiment extending between 1982--1986 at Brookhaven National Laboratory (BNL) by Alburger et al.\ \cite{Alburger} on the half-life of $^{32}$Si exhibited clear annual periodicities, which the authors could not account for in terms of temperature variations or other conventional environmental influences \cite{Javorsek_intro}. Similarly, data taken between 1983 and 1998 at the Physikalisch-Technische Bundesanstalt (PTB) in Germany \cite{PTB} exhibited annual variations similar to those observed at BNL. Some have argued that such fluctuations are due to experimental influences and improper uncertainty calculations \cite{Pomme}.

A possible interpretation of these data is that they may be attributable to the annually varying flux of solar neutrinos, or from a contribution of axionic dark matter, which were perturbing nuclear decay rates. Support for this conjecture comes from a report by Davis \cite{Davis} of an increase in the number of solar neutrinos detected in his Homestake solar neutrino experiment associated with a solar storm on 4 June 1991. The subsequent observation of a precipitous drop in the decay rate of $^{54}$Mn associated with a significant solar storm on 13 December 2006 \cite{flare}, further supports the suggestion that at least some nuclear decay rates could be influenced by changes in the local flux of neutrinos, or other particles such as axionic dark matter coupling to baryons.

Notwithstanding the implications of the preceding discussion, it has also become clear that not all nuclei exhibit fluctuations in response to the same perturbations. This follows, for example, from the discussion in Ref.~\cite{inspiral2} of data acquired during the GW170817 neutron star inspiral: of the three isotopes studied in the experiment described there, $^{44}$Ti and $^{60}$Co exhibited an effect similar to that observed in Ref.~\cite{inspiral1} for $^{32}$Si and $^{36}$Cl, but $^{137}$Cs did not. Since at present there is no theory to account for how some nuclear decay rates could be influenced by a flux of neutrinos or perhaps other solar influences, it is not surprising that some nuclei may be more responsive to an external influence than others. 

A recent paper provides additional support for a theory in which the observed periodic variations in radioactive decays could arise from dark matter coupling to baryonic number \cite{Agafonova}.  This reference explores the similarity between signals seen in neutrino detectors and gravitational antennas during SN1987A, and a similar signal during the GW178017 neutron star inspiral in an experiment monitoring the half-life of $^{32}$Si.  It is demonstrated that the similarity of these signals is attributed to the influence of a gravity wave on a component of dark matter that couples to baryons.  Since this component could directly influence the decay rates of unstable nuclei, this picture provides a natural mechanism to account for gravitationally-induced variations in radioactive decays.

Although there is at present no theory to explain how different nuclei might respond differently to changes in the local flux of neutrinos, axions, or other particles in the interstellar medium, one possible mechanism relies on the known sensitivity of nuclear decay rates to the available phase space for their daughter particles. 

One approach that is being explored is to assume that a component of the ambient stellar medium influences some nuclear decay rates by modifying the phase space available for their daughter particles. This ``medium'' approach has two attractive features: First, it could explain the surprising observation \cite{Mueterthies} that the annual fluctuations $\Delta \Gamma_{i}/\Gamma_{i}$ in many decay rates $\Gamma_{i}$ are approximately the same, even though the magnitudes of the $\Gamma_{i}$ vary over 9 orders of magnitude. Secondly, the neutrino kinematics in the medium picture allow for the effective observed neutrino mass $(m_{\nu}^{2})_{\rm eff}$ to be negative, in agreement with observation, even though the intrinsic neutrino mass ($m_{\nu}^{2}$) is positive \cite{Mueterthies}.

In summary, there is at present a growing literature from well-done experiments suggesting that some nuclear decay rates may exhibit fluctuations arising from external influences. There is also an extensive literature from equally well-done experiments where such fluctuations were sought but not found. Since at present, there is no quantitative theory to account for the observed fluctuations, it is difficult to draw any firm conclusions about the origins of the observed fluctuations. It is clear that much remains to be learned about potential external influences on radioactive decay rates, from both  experimental and theoretical standpoints. 

As a means to organize the literature we have noted for each reference whether they have reported variations (V) or no variations (NV) from the expected exponential decay law. We have separately considered $\alpha$-decays, $\beta$-decays, electron capture-decays and $\gamma$-decays, where appropriate. The present situation is further complicated by disagreements in the published values of the half-lives of various radionuclides, some of which we present in an appendix to the main review. Although these disagreements can be reconciled using various algorithms, the question remains as to whether they could arise from the same mechanisms which constitute the body of this review. Given the possible connection of this literature to the central focus of the present review, we have also included a selection of appropriate references, which are denoted by ``D''. 

In the bibliography, we have organized the literature alphabetically to allow for easy access to specific papers. Additionally, in Table \ref{longtable} we have cross-referenced the experiments in a manner that allows the interested reader to focus on common features of specific experiments, such as which isotopes were studied and which detectors were employed.

By organizing the literature in this way, we hope to aid researchers in resolving the many issues that remain in this field.  Any attempt to explain the anomalies by experimental or environmental effects will need to reproduce the observations in a convincing way.  Similarly, any theory requiring beyond the Standard Model physics to explain the anomalies needs to be consistent with all other observations and be testable in other types of experiments.

Finally, we recognize that despite our many efforts to compile a useful reference which is as complete as possible, we may have inadvertently omitted some references which should have been included. Since we plan to update this bibliography as needed, we invite suggestions for additional references to be included in future editions. These suggestions should be sent to the corresponding author. 

\ack This article is based upon work sponsored by the Defense Advanced Research Projects Agency (DARPA). Any opinions, findings, conclusions or recommendations ex-pressed are those of the authors and do not necessarily reflect the views of DARPA, the U.S. Air Force, the U.S. Department of Defense, or the U.S. Government. We wish to thank Bianca Caminada, Emily Kincaid, Claire Landgraf, Connor Mohs, Connor Petway, and Ethan Zweig for their help in compiling this bibliography.

{\tiny
\begin{longtable}{|l|l|l|l|p{2.5cm}|p{2.7cm}|p{2cm}|}
\caption{Partial list of time varying and time-independent decay experiments. Institution abbreviations are as listed: GSL: Gran Sasso Laboratory; GSIL: Geological Survey of Israel Laboratory; OPC: Optical Physics Company  MSU: Moscow State University; BNL: Brookhaven National Laboratory; PTB: Physikalisch-Technische Bundesanstalt ; LMSU The Lomonosov Moscow State University; Baylor: Baylor College of Medicine; U.S.A.F. Academy: U.S. Air Force Academy; KIT: Karlsruhe Institute of Technology; Karpov Institute: Karpov Institute of Physical Chemistry;CRIM: Central Research Institute of Machine, IIRES: Institute for Industrial, Radiophysical and Environmental Safety}
\label{longtable} 
\\ \hline
\textbf{Isotope}                                      & \textbf{Decay}         		& \textbf{Detector Type}       & \textbf{Detected}   		& \textbf{Observations} 			&\textbf{Institution}               			& \textbf{Reference}          \\ \hline
\endfirsthead
\mbox{} \\ \hline
\textbf{Isotope}                                      & \textbf{Decay}         		& \textbf{Detector Type}       & \textbf{Detected}   		& \textbf{Observations} 			&\textbf{Institution}               			& \textbf{Reference}          \\ \hline
\endhead
\hline \multicolumn{7}{r}{Continued on next page}
\endfoot
\hline
\endlastfoot
\textsuperscript{3}H                          & $\beta^{-}$          & Liquid Scintillator    & $\beta^{-}$           &   1yr, 12.1yr, 18d, 42d, 12.51yr                     & Novi Sad, Purdue, USAFA, Karpov, MSU      &   \cite{Bikit}, \cite{Kay}, \cite{LeePaper}, \cite{Veprev}, \cite{Shnoll}     \\ \hline 
\textsuperscript{3}H                          & $\beta^{-}$          & Photodiodes   & $\beta^{-}$           &    1yr                    & Purdue, Uhldingen, OPC,       &       \cite{Kay} ,  \cite{Falkenberg} , \cite{Jenkins_arXiv},           \\ \hline 
\textsuperscript{3}H                          & $\beta^{-}$          & Solid State   & $\beta^{-}$           &   2yr                   & Purdue, KIT      &       \cite{Jenkins_arXiv} ,  \cite{Lobashev}            \\ \hline 
\textsuperscript{14}C                         & $\beta^{-}$          & Liquid Scintillator   & $\beta$           &  No effect                   & Khalifa, USAFA      &   \cite{Goddard}, \cite{LeePaper}            \\ \hline 
\textsuperscript{18}F                         & $\beta^{+}$          & Ion Chamber    & $\gamma$        &    No effect                   & PTB      &   \cite{Schrader_F18}            \\ \hline 
\textsuperscript{22}Na 			& $\beta^{+}$ 		 & Solid State (Ge)    & $\gamma$             & 1yr                    & Berkeley				       & \cite{O'Keefe}                  \\ \hline
\textsuperscript{22}Na                        & $\beta^{+}$           & HPGe                &    $\gamma$                   & No effect                           &     Novi Sad , Berkeley                  & \cite{Knezevic} , \cite{Norman}                     \\ \hline
\textsuperscript{22}Na                        & $\beta^{+}$           & Geiger M\"{u}ller   &    $\beta^{-}$                  & No effect                          &      BYU,                                  & \cite{McKnight}, \cite{Bergeson}, \cite{Ware}               \\ \hline
\textsuperscript{32}Si                       & $\beta^{-}$          & Scintillation         & $\gamma$             & GW inspiral, 1yr    & Purdue, BNL              &  \cite{GW} , \cite{Alburger_Si}                 \\ \hline
\textsuperscript{32}Si                       & $\beta^{-}$          & Ge(Li)         & $\gamma$             & 1yr   & CRIM              &  \cite{Baurov_Si32}                  \\ \hline
\textsuperscript{32}Si                        & $\beta^{-}$          & Proportional        &  $\beta^{-}$                  & 1yr                      & BNL                                    &    \cite{Alburger}, \cite{Heim}, \cite{JenkinsSi32}, \cite{Sturrock BNL}               \\ \hline
\textsuperscript{32}Si/\textsuperscript{36}Cl                        & $\beta^{-}$          & Proportional        &  $\beta^{-}$                  & No effect                       &  Wadworth Center                                   &   \cite{Semkow}   \\ \hline
\textsuperscript{32}Si/\textsuperscript{36}Cl                        & $\beta^{-}$          & Ion Chamber       &  $\gamma$        & 27d,  1yr,             &  PTB    &   \cite{Sturrock},  \cite{Sturrock_PTB}  \\ \hline
\textsuperscript{36}Cl                        & $\beta^{-}$          & Proportional        & $\beta^{-}$           & 1yr, 11.71yr, 2.11yr	  & Purdue, BNL         & \cite{Kay},  \cite{MohsinallyThesis}, \cite{Jenkins_arXiv}, \\ \hline
\textsuperscript{36}Cl                      & $\beta^{-}$          & Scintillation         & $\gamma$             & GW inspiral     & Purdue              &  \cite{GW}                  \\ \hline
\textsuperscript{36}Cl                      & $\beta^{-}$          & Scintillation         & $\gamma$             & No effect     & PT B              &  \cite{KossertCl36}                  \\ \hline
\textsuperscript{36}Cl                        & $\beta^{-}$          & Geiger M\"{u}ller     & $\beta^{-}$           & 1yr                       & Purdue                      & \cite{Jenkins_Astroparticle}, \cite{Kay}, \cite{MohsinallyThesis}               \\ \hline
\textsuperscript{36}Cl                        & $\beta^{-}$          & Geiger M\"{u}ller     & $\beta^{-}$           & No effect                & BYU                         & \cite{Bergeson}            \\ \hline
\textsuperscript{40}K                         & $\beta^{-}$, EC       & NaI Crystal         &    $\gamma$                     & No effect                            &     TBD                                   &  \cite{Bellotti2013}, \cite{Bellotti2015}, \cite{Bellotti2018}                  \\ \hline
\textsuperscript{44}Ti                        & EC                  & NaI(TI)             &    $\gamma$                      & No effect                             &      Zurich, Amsterdam                                  &   \cite{Breur}  , \cite{Angevaare}               \\ \hline
\textsuperscript{44}Ti                        & EC                  & HPGe                &    $\gamma$                   & No effect                          &     Berkeley                                   &  \cite{Norman}                  \\ \hline
\textsuperscript{54}Mn                        & EC           & Scintillation       & $\gamma$             & Solar flare                 & Purdue                      & \cite{Jenkins_solar_flare}               \\ \hline
\textsuperscript{54}Mn                        & EC           & Scintillation       & $\gamma$             & 1yr                       & Purdue, Baylor                      & \cite{Jenkins_arXiv}, \cite{Ellis}               \\ \hline
\textsuperscript{56}Mn                        & EC           & Scintillation       & $\gamma$             & 1yr                       & Purdue                      & \cite{Jenkins_arXiv}              \\ \hline
\textsuperscript{55}Fe                        & EC                  & Scintillation                  &      $\gamma$                & No effect                            &    PTB                                    &   \cite{KossertFe}                 \\ \hline
\textsuperscript{60}Co                        & $\beta^{+}$    & NaI(TI)             &    $\gamma$                  & No effect                             &  Zurich, Amsterdam                                   & \cite{Breur}, \cite{Angevaare}                    \\ \hline
\textsuperscript{60}Co                        & $\beta^{+}$    & NaI(TI)             &    $\gamma$                  & 1d, 27d, 1yr           &  CRIM                                   &   \cite{Baurov_Co60}, \cite{Baurov_Si32}                      \\ \hline
\textsuperscript{60}Co                        & $\beta^{+}$    & Scintillation      & 	$\gamma$             & 1d, 12.11yr, 10d, 20d, 27d            &    CRIM                                 &\cite{Baurov}, \cite{Baurov_arXiv}            \\ \hline
\textsuperscript{60}Co                        & $\beta^{+}$          & HPGe     & $\gamma$ 			& 1yr                     			  & IMS             		& \cite{Lee}          \\ \hline
\textsuperscript{56}Co                        & $\beta^{+}$        & Ge(Li)     & $\gamma$ 			&  No effect                       & BNL             & \cite{Alburger_Co56}         \\ \hline
\textsuperscript{60}Co                        & $\beta^{+}$     & Geiger M\"{u}ller     & $\beta^{-}$  		& 1yr                      & LMSU             & \cite{Parkhomov_arXiv},\cite{Parkhomov_alpha_beta}          \\ \hline
\textsuperscript{60}Co                        & $\beta^{+}$     & Geiger M\"{u}ller     & $\beta^{-}$		 & No effect                & BYU                        & \cite{Bergeson} 	 \\ \hline
\textsuperscript{65}Zn                        & EC    & Ion Chamber     & $\gamma$	& No effect                & PTB                    & \cite{Schrader_F18} 	 \\ \hline
\textsuperscript{67}Ga                        & EC    & Ion Chamber     & $\gamma$	& No effect                & PTB                    & \cite{Schrader_F18} 	 \\ \hline
\textsuperscript{81}Rb                        & EC   & Ion Chamber     & $\gamma$	& No effect                & PTB                    & \cite{Schrader_F18} 	 \\ \hline
\textsuperscript{82}Rb                        & EC    & Ion Chamber     & $\gamma$	& No effect                & PTB                    & \cite{Schrader_F18} 	 \\ \hline
\textsuperscript{85}Kr                        & $\beta^{-} $  & Ion Chamber     & $\gamma$	& 1yr		               & PTB                    & \cite{Sturrock_AstrophysJour}, \cite{Kay}, \cite{MohsinallyThesis}, \cite{Schrader_2016}	 \\ \hline
\textsuperscript{85}Kr                        & $\beta^{-} $    & Ion Chamber     & $\gamma$	& No effect                & PTB                    & \cite{Schrader_F18} 	 \\ \hline
\textsuperscript{90}Sr 		& $\beta^{-}$   	& Geiger M\"{u}ller     & $\beta^{-}$    & 1y, 11.71yr, 10yr    & LMSU, Purdue   & \cite{Parkhomov_arXiv},\cite{Parkhomov_alpha_beta}, \cite{Jenkins_arXiv}, \cite{MohsinallyThesis}      \\ \hline
\textsuperscript{90}Sr 		& $\beta^{-}$           & Geiger M\"{u}ller    &   $\beta^{-}$     & No effect                &      PTB, Khalifa       &  \cite{McKnight}, \cite{Kossert_Sr}, \cite{Goddard}                  \\ \hline
\textsuperscript{90}Sr  		& $\beta^{-}$           & Scintillation  		&   $\beta^{-}$     & 10yr                          &      PTB       & \cite{Kossert_Sr}                  \\ \hline
\textsuperscript{90}Sr  		 & $\beta^{-} $    		& Ion Chamber     & $\gamma$		& No effect                & PTB                    & \cite{Schrader_F18} 	 \\ \hline
\textsuperscript{99}Tc   		& $\beta^{-} $    		& Ion Chamber     & $\gamma$		& No effect                & PTB                    & \cite{Schrader_Eu}, \cite{Schrader_Ra} 	 \\ \hline
\textsuperscript{99}Mo   		& $\beta^{-} $    		& Ion Chamber     & $\gamma$		& No effect                & PTB                    & \cite{Schrader_F18} 	 \\ \hline
\textsuperscript{106}Ru  		 & $\beta^{-} $    		& Ion Chamber     & $\gamma$		& No effect                & PTB                    & \cite{Schrader_F18} 	 \\ \hline
\textsuperscript{108}Ag                       & EC            & Ion Chamber         & $\gamma$             & 1yr                     & PTB  & \cite{Kay}, \cite{MohsinallyThesis}              \\ \hline
\textsuperscript{108}Ag                       & EC            & Ion Chamber         & $\gamma$             & No effect                     & PTB  & \cite{Schrader},  \cite{Schrader_2016},               \\ \hline
\textsuperscript{108}Ag                       & EC            & HPGe         	& $\gamma$             & No effect                     & Berkeley   &  \cite{Norman}         \\ \hline
\textsuperscript{109}Cd                       & EC            & Ion Chamber         & $\gamma$             & No effect                     & PTB  & \cite{Schrader_F18}              \\ \hline
\textsuperscript{111}In                       & EC            & Ion Chamber         & $\gamma$             & No effect                     & PTB  & \cite{Schrader_F18}              \\ \hline
\textsuperscript{121}Sn                       & $\beta^{-} $               & HPGe         	& $\gamma$             & No effect                       & Berkeley  & \cite{Norman}              \\ \hline
\textsuperscript{123}I                        & EC            & Ion Chamber         & $\gamma$             & No effect                     & PTB  & \cite{Schrader_F18}              \\ \hline
\textsuperscript{131}I                        & EC            & Ion Chamber         & $\gamma$             & No effect                     & PTB  & \cite{Schrader_F18}              \\ \hline
\textsuperscript{133}Ba                       & EC            & Ion Chamber         & $\gamma$             & 1yr                & Purdue  & \cite{Kay}, \cite{MohsinallyThesis},          \\ \hline
\textsuperscript{133}Ba                       & EC            & Ion Chamber         & $\gamma$             & No effect                     & PTB  & \cite{Schrader}, \cite{Schrader_2016}         \\ \hline
\textsuperscript{133}Ba                       & EC           & HPGe                  &   TBD                   & No effect                             &      Berkeley                                  & \cite{Norman}                   \\ \hline
\textsuperscript{137}Cs                       & $\beta^{-}$          & Scintillation       & $\gamma$             & 1d, 12.11yr             & CRIM,  Purdue, OSU                                   & \cite{Baurov}, \cite{Buncher}, \cite{Lee}, \cite{MohsinallyThesis},\cite{Jenkins_arXiv}        \\ \hline
\textsuperscript{137}Cs                       & $\beta^{-}$           & NaI(TI)             &    $\gamma$                      & No effect                            &    Zurich                                    &   \cite{Breur}                 \\ \hline
\textsuperscript{137}Cs                       & $\beta^{-}$          &  Geiger M\"{u}ller              &   $\beta^{-}$                      & No effect                            &    PTB                                    &    \cite{McKnight}                \\ \hline
\textsuperscript{137}Cs                       & $\beta^{-}$          &  Geiger M\"{u}ller              &   $\beta^{-}$                      & Solar Eclipse                         &    Greenland                                    &    \cite{Javorsek}                \\ \hline
\textsuperscript{137}Cs                       & $\beta^{-}$          & HPGe                &   $\gamma$                    & No effect                            &   TBD, IMS                                 & \cite{Bellotti_Cs} , \cite{Lee}                             \\ \hline
\textsuperscript{137}Cs                       & $\beta^{-}$             & Ion Chamber         & $\gamma$             & 27d                    & PTB  & \cite{Sturrock_AstrophysJour},       \\ \hline
\textsuperscript{137}Cs                       & $\beta^{-}$             & Ion Chamber         & $\gamma$             & No effect                     & PTB  & \cite{Schrader},       \\ \hline
\textsuperscript{152}Eu                       & EC, $\beta^{-}$    & Solid State (Ge)    & $\gamma$             & 1yr                      & Purdue, PTB, USAFA                                    & \cite{JenkinsSi32}, \cite{Kay}, \cite{Siegert}, \cite{Lee}              \\ \hline
\textsuperscript{152}Eu                       & EC, $\beta^{-}$ 	& HPGe & $\gamma$             & 1yr                       & IMS                                    & \cite{Lee}              \\ \hline
\textsuperscript{152}Eu                       & EC, $\beta^{-}$ 	& Ion Chamber         & $\gamma$             & 1yr                    & PTB, USAFA  & \cite{Kay}, \cite{Lee}             \\ \hline
\textsuperscript{152}Eu                       & EC, $\beta^{-}$	 & Ion Chamber         & $\gamma$             & No effect                       &  PTB & \cite{Schrader}, \cite{Schrader_2016}             \\ \hline
\textsuperscript{153}Sm                       & $\beta^{-}$	 & Ion Chamber         & $\gamma$             & No effect                       &  PTB & \cite{Schrader}             \\ \hline
\textsuperscript{154}Eu                       & $\beta^{-}$	 & Ion Chamber         & $\gamma$             &  1yr                     &  Purdue & \cite{Kay} , \cite{MohsinallyThesis},         \\ \hline
\textsuperscript{154}Eu                       & $\beta^{-}$	 & Ion Chamber         & $\gamma$             &  No effect                    &  PTB & \cite{Schrader_F18} , \cite{Schrader_Eu},  \cite{Schrader_2016}               \\ \hline
\textsuperscript{169}Er                       & $\beta^{-}$	 & Ion Chamber         & $\gamma$             &  No effect                    &  PTB & \cite{Schrader_F18}           \\ \hline
\textsuperscript{177}Lu                       & $\beta^{-}$	 & Ion Chamber         & $\gamma$             &  No effect                    &  PTB & \cite{Schrader_F18}           \\ \hline
\textsuperscript{186}Re                      & $\beta^{-}$	 & Ion Chamber         & $\gamma$             &  No effect                    &  PTB & \cite{Schrader_F18}           \\ \hline
\textsuperscript{198}Au                       & $\beta^{-}$           & HPGe                &   $\gamma$                    & Neutrino flux                         & NIST                                   &  \cite{LindstromAu}, \cite{Lindstrom}            \\ \hline
\textsuperscript{198}Au                       & $\beta^{-}$           & HPGe                &   $\gamma$                    & No effect                            &   Texas A\&M                                     &  \cite{Hardy}, \cite{Goddard_Au198}                  \\ \hline
\textsuperscript{201}Ti                       & $\beta^{-}$           & Ion Chamber                &   $\gamma$                    & No effect                            &   PTB                                     &  \cite{Schrader_F18}                  \\ \hline
\textsuperscript{204}Ti                       & $\beta^{-}$           & Geiger M\"{u}ller                &   $\gamma$                    & No effect                            &   Khalifa                                     &  \cite{Goddard}                  \\ \hline
\textsuperscript{222}Rn                       & $\alpha$, $\beta^{-}$& Scintillation       & $\gamma$             & 1yr, 11.71yr, 2.11yr, 1d  & GSIL, PTB 		& \cite{Steinitz}, \cite{Sturrock_astroparticle}, \cite{Jenkins_arXiv}      \\ \hline
\textsuperscript{222}Rn                       & $\alpha$, $\beta^{-}$& NaI       		& $\gamma$             & 1d, 27d, 1yr  			& GSIL 		&\cite{Steinitz}, \cite{Steinitz_ESS}      \\ \hline
\textsuperscript{222}Rn                       & $\alpha$, $\beta^{-}$& Ion Chamber       & $\gamma$             & No effect  			& PTB 		& \cite{Schrader_F18}      \\ \hline
\textsuperscript{224}Rn                       & $\alpha$, $\beta^{-}$& Ion Chamber       & $\gamma$             & No effect  			& PTB 		& \cite{Schrader_F18}      \\ \hline
\textsuperscript{226}Ra                       & $\alpha$, $\beta^{-}$& Ion Chamber         & $\gamma$             & 1yr, 11.71yr, 2.11yr  	& Purdue, BNL, Wadworth Center         & \cite{Jenkins_arXiv}, \cite{Kay}, \cite{MohsinallyThesis}, \cite{Sturrock_PTB}  \cite{Sturrock_Solar_Physics}, \cite{Schrader_Eu}, \cite{Semkow} \\ \hline
\textsuperscript{226}Ra                       & $\alpha$, $\beta^{-}$& Ion Chamber         & $\gamma$             & No effect  	& PTB        & \cite{Nahle}, \cite{Schrader_Ra}, \cite{Schrader_2016} \\ \hline
\textsuperscript{226}Ra                       &$\alpha$, $\beta^{-}$          & Geiger M\"{u}ller    &   $\gamma$                    & No effect                            &   IIRES                                     &  \cite{Sanchez}                  \\ \hline
\textsuperscript{226}Ra                      & $\alpha$, $\beta^{-}$            & NaI Crystal         &  $\gamma$                        & 1d                           &   GSIL                                     &  \cite{Steinitz_Ra226}                   \\ \hline
\textsuperscript{232}Th                       & $\alpha$            & NaI Crystal         &  $\gamma$                        & No effect                            &   GSL                                     &  \cite{Bellotti2013} , \cite{Bellotti2015}                  \\ \hline
\textsuperscript{238}Pu                       & $\beta^{-}$          & SpaceCraft         & $\alpha$             & Seasonal   & Wabash, Purdue              &  \cite{Krause}                 \\ \hline
\textsuperscript{239}Pu                       & $\beta^{-}$          & Solid State         & $\alpha$             & 1d, 1yr, 13.51yr    & MSU, Purdue              &  \cite{Shnoll} , \cite{Kay}, \cite{ MohsinallyThesis}                \\ \hline
\textsuperscript{239}Pu                       & $\beta^{-}$          & Geiger M\"{u}ller    &$\gamma$        & No effect                            &   LMSU                                     &  \cite{Sanchez}                  \\ \hline

\hline

\end{longtable}
}


\pagebreak

\begin{flushleft}
\large\textbf{Key to Paper Categories}
\end{flushleft}

\begin{flushleft}
General Categories    \qquad \qquad \qquad Qualifiers      
\end{flushleft}

\begin{table}[H]
\begin{tabular}{lllllll}
V                  & Variations    &  &  &  & E          & Experimental     \\
NV                 & No Variations &  &  &  & P          & Phenomenological \\
A                  & Alpha Decay   &  &  &  & T          & Theoretical     \\
B                 & Beta Decay &  &  &  & L          & Laboratory \\
EC                 & Electron Capture  &  &  &  & R          & Review \\
G                 & Gamma Decay  &  &  &  & S          & Solar Influence \\
D		& Disagreements
\end{tabular}
\end{table}

\begin{flushleft}
\large\bf Bibliography
\end{flushleft}

\begin{justify}
The following is a list of references which directly discuss the implication of time-dependent variations (V) in radioactive decay rate data as well as arguments against such variations (NV).
\end{justify}

\begin{enumerate}
\renewcommand{\labelenumi}{\arabic{enumi}.}

\bibitem{Alburger_Si} Alburger D. E., Harbottle G., Norton E. F., \href{https://doi.org/10.1016/0012-821X(86)90058-0}{``Half-life of \textsuperscript{32}Si,''  \textit{Earth and Planetary Science Letters,}} 1986, {\bf 78}, Iss. 2-3, 168-176. \textbf{Topics: B,D,E,V}

\bibitem{Alburger_Co56} Alburger D. E., Wesselborg C., \href{https://www.osti.gov/biblio/6482874}{``Precision measurement of the half-life of \textsuperscript{56}Co,''  \textit{International Symposium on Capture Gamma-ray Spectroscopy and Related Topics,}} 1990. \textbf{Topics: B,D,E,NV}

\item Aldrich L. T., Nier A. O., \href{https://doi.org/10.1103/PhysRev.74.876}{``Argon 40 in Potassium Minerals,''  \textit{Physical Review,}} 1948, {\bf 74}, No. 8, 876-877. \textbf{Topics: B,E,V}

\item Alexeev E. N., Gangapshev A. M.,  Gavrilyuk Y. M., Gezhaev A. M., Kazalov V. V., Kuzminov V. V., Panasenko S. I., Petrenko O. D., Ratkevich S. S., ``Annula variations of the \textsuperscript{214}Po, \textsuperscript{213}Po and \textsuperscript{212}Po half-life values,'' \href{https://arxiv.org/abs/2010.08283v1}{\underline{arXiv:2010.08283}} [nucl-ex], 2020, \textbf{Topics: A,E,P,V}

\item Alexeev E. N., Gavrilyuk Y. M., Gangapshev A. M., Gezhaev A. M., Kazalov V. V., Kuzminov V. V., Panasenko S. I., Ratkevich S. S., \href{https://doi.org/10.1134/S1063779618040044}{``Search for Variatons of \textsuperscript{213}Po Half-Life,''  \textit{Physics of Particles and Nuclei,}} 2018, {\bf 49}, No. 4, 557-562. \textbf{Topics: A,E,V}

\item Alexeyev E. N., Alekseenko V. V., Gavriljuk J. M., Gangapshev A. M., Gezhaev A. M., Kazalov V. V., Kuzminov V. V., Panasenko S. I., Ratkevich S. S., Yakimenko S. P., \href{https://doi.org/10.1016/j.astropartphys.2013.04.005}{``Experimental test of the time stability of the half-life of alpha-decay \textsuperscript{214}Po nuclei,''  \textit{Astroparticle Physics,}} 2013, {\bf 46}, 23-28. \textbf{Topics: A,E,V}

\item Alexeyev E. N., Gavrilyuk Y. M., Gangapshev A. M., Kazalov V. V., Kuzminov V. V., Panasenko S. I., Ratkevich S. S., \href{https://doi.org/10.1134/S1063779616060034}{``Results of a Search for Daily and Annual Variations of the \textsuperscript{214}Po Half-Life at the Two Year Observation Period,''  \textit{Physics of Particles and Nuclei,}} 2016, {\bf 47}, No. 6, 986-994. \textbf{Topics: A,E,V}

\item Alexeyev E. N., Gavrilyuk Y. M., Gangapshev A. M., Kazalov V. V., Kuzminov V. V, Panasenko S. I., Ratkevich S. S., \href{https://doi.org/10.1134/S1063779615020021}{``Sources of the Systematic Errors in Measurements of \textsuperscript{214}Po Decay Half-Life Time Variations at the Baksan Deep Underground Experiments,''  \textit{Physics of Particles and Nuclei,}} 2015, {\bf 46}, 157-165. \textbf{Topics: A,E,P,V}

\bibitem{Angevaare} Angevaare J. R., Barrow P., Baudis L., Breur P. A., Brown A.,  Colijn A. P., Cox G., Gienal M., Gjaltema F., Helmling-Cornell A., et al., \href{https://iopscience-iop-org.ezproxy.lib.purdue.edu/article/10.1088/1748-0221/13/07/P07011}{``A precision experiment to investigate long-lived radioactive decays,''  \textit{Journal of Instrumentation,}} 2018, {\bf 13}. \textbf{Topics: A,B,E,EC,G,NV}

\item Bahcall J. N., Press W. H., \href{http://adsabs.harvard.edu/full/1991ApJ...370..730B}{`Solar-cycle modulation of event rates in the chlorine solar neutrino experiment,''  \textit{Astrophysical Journal,}} 1991, {\bf 370}, 730-742. \textbf{Topics: E,S,V}

\item Bahcall J. N., Field G. B., Press W. H., \href{http://articles.adsabs.harvard.edu/full/1987ApJ...320L..69B}{`Is solar neutrino capture rate correlated with sunspot number?,''  \textit{Astrophysical Journal,}} 1987, {\bf 320}, L69-L73. \textbf{Topics: E,S,V}

\item Bahcall J. N., Serenelli A. M., Basu S., \href{https://iopscience.iop.org/article/10.1086/428929/meta}{`New Solar Opacities, Abundances, Helioseismology, and Neutrino Fluxes,''  \textit{Astrophysical Journal Letters,}} 2005, {\bf 621}, No. 1, L85-L88. \textbf{Topics: E,S,V}

\item Barnes V. E., Bernstein D. J., Bryan C. D., Cinko N., Deichert G. G., Gruenwald J. T., Heim J. M., Kaplan H. B., LaZur R., Neff D., et al. ``Search for Perturbations of Nuclear Decay Rates Induced by Reactor Electron Antineutrinos,''  \href{https://arxiv.org/abs/1606.09325} {\underline{arXiv:1606.09325}} [nucl-ex], 2016. \textbf{Topics: B,E,EC,P,S,V} 

\item Barnes V. E., Bernstein D. J., Bryan C. D., Cinko N., Deichert G. G., Gruenwald J. T., Heim J. M., Kaplan H. B., LaZur R., Neff D., et al. \href{https://doi.org/10.1016/j.apradiso.2019.01.027}{``Upper limits on perturbations of nuclear decay rates induced by reactor electron antineutrinos,''  \textit{Applied Radiation and Isotopes,}} 2019, {\bf 149}, 182-199. \textbf{Topics: B,E,T,S,V}

\item Barnett S. M., Huttner B., Loudon R., Matloob R., \href{https://iopscience.iop.org/article/10.1088/0953-4075/29/16/019}{``Decay of excited atoms in absorbing dielectrics,''  \textit{Journal of Physics B: Atomic, Molecular and Optical Physics,}} 1996, {\bf 29}, No. 16. \textbf{Topics: T,V}

\item Bashindzhagyan G., Barnes V., Fischbach E., Hovsepyan G., Korotkova N., Merkin M., Sinev N., Voronin A., \href{https://pos.sissa.it/236/136/pdf}{``Solar Influence on Decay Rate (SIDR) Experiment,''  \textit{The 34th International Cosmic Ray Conference,}} 2015, {\bf 236}. \textbf{Topics: B,E,T,V}

\item Bashindzhagyan G., Barnes V., Fischbach E., Hovsepyan G., Korotkova N., Poghosyan L., Sinev N., \href{https://doi.org/10.22323/1.301.0050}{``SIDR experiment status and first results,''  \textit{The 35th International Cosmic Ray Conference,}} 2017, {\bf 301}. \textbf{Topics: B,E,T,V}

\item Baurov Y. A., \href{https://doi.org/10.3103/S1062873812040065}{``The Anisotropic Phenomenon in the \textbeta Decay of Radioactive Elements and in Other Processes in Nature,'' \textit{Bulletin of the Russian Academy of Sciences: Physics,}} 2012, {\bf 76}, No. 10, 1076-1080. \textbf{Topics: B,E,V}

\item Baurov Y. A., Albanese L., Meneguzzo F., Menshikov V. A., \href{https://www.researchgate.net/publication/260818491_Protecting_the_Planet_from_the_Asteroid_Hazard}{``Protecting the planet from the asteroid hazard,''  \textit{International Journal of Pure and Applied Physics,}} 2013, {\bf 9} 151-168. \textbf{Topics: B,P,S,T,V}

\bibitem{Baurov} Baurov Y. A., Konradov A. A., Kushniruk V. F., Kuznetsov E. A., Sobolev Y. G., Ryabov Y. V., Senkevich A. P., Zadorozsny S. V., \href{https://doi.org/10.1142/S0217732301005187}{``Experimental investigations of changes in \textbeta-decay rate of \textsuperscript{60}Co and \textsuperscript{137}Cs,''  \textit{Modern Physics Letters A,}} 2001, {\bf 16} No. 32, 2089-2101. \textbf{Topics: E,P,S,V}

\bibitem{Baurov_arXiv} Baurov Y. A., Nikitin V. A., Dunin V. B., Demchuk N. A., Baurov  A. Y., Tihomirov V. V., Sergeev S. V., Baurov A. Y. Jr., ``Results of experimental investigations of \textsuperscript{60}Co \textbeta -decay rate variation,''  \href{https://arxiv.org/abs/1304.6885}{\underline{arXiv:1304.6885}} [nucl-ex], 2013. \textbf{Topics: B,E,V}

\item Baurov Y. A., Malov I. F., ``Variations of decay rates of radio-active elements and their connections with global anisotropy of physical space,''  \href{https://arxiv.org/abs/1001.5383v1}{\underline{arXiv:1001.5383}} [physics.gen-ph], 2010. \textbf{Topics: A,B,E,S,V}

\bibitem{Baurov_Co60} Baurov Y. A., Sobolev Y. G., Ryabov Y. V., Kushniruk V. F.,  \href{https://doi.org/10.1134/S1063778807110014}{``Experimental Investigations of Changes in the Rate of Beta Decay of Radioactive Elements,''  \textit{Physics of Atomic Nuclei,}} 2007, {\bf 70}, 1825-1835. \textbf{Topics: B,E,V}

\bibitem{Baurov_Si32} Baurov Y. A., Sobolev Y. G., Kushniruk V. F., Kuznetzov E. A., Konradov A. A., \href{https://arxiv.org/abs/hep-ex/9907008v1}{``Experimental investigation of changes in \textbeta-decay count rate of radioactive elements,'' \underline{arXiv:hep-ex/9907008}}, 1999. \textbf{Topics: B,E,V}

\bibitem{BellottiRn} Bellotti E., Broggini C., Di Carlo G., Laubenstein M., Menegazzo R., \href{https://doi.org/10.1016/j.physletb.2015.03.021}{``Precise measurement of \textsuperscript{222}Rn half-life: A probe to monitor the stability of radioactivity,''  \textit{Physics Letters B,}} 2015, {\bf 743}, 526-530. \textbf{Topics: A,B,E,NV,P}

\bibitem{Bellotti2013} Bellotti E., Broggini C., Di Carlo G., Laubenstein M., Menegazzo R., \href{https://doi.org/10.1016/j.physletb.2013.02.002}{``Search for correlations between solar flares and decay rate of radioactive nuclei,''  \textit{Physics Letters B,}} 2013, {\bf 710}, Iss. 1-3, 116-119. \textbf{Topics: A,B,E,EC,NV,P,S,}

\bibitem{Bellotti_Cs} Bellotti E., Broggini C., Di Carlo G., Laubenstein M., Menegazzo R., \href{https://doi.org/10.1016/j.physletb.2012.02.083}{``Search for time dependence of the \textsuperscript{137}Cs decay constant,''  \textit{Physics Letters B,}} 2012, {\bf 710}, Iss. 1, 114-117. \textbf{Topics: B,E,NV}

\bibitem{Bellotti2018} Bellotti E., Broggini C., Di Carlo G, Laubenstein M., Menegazzo R., \href{https://doi.org/10.1016/j.physletb.2018.02.065}{``Search for time modulations in the decay rate of \textsuperscript{40}K and \textsuperscript{226}Ra at the underground Gran Sasso Laboratory,''  \textit{Physics Letters B,}} 2018, {\bf 780}, 61-65. \textbf{Topics: A,E,NV,P,S}

\item Bellotti E., Broggini C., Di Carlo G., Laubenstein M., Menegazzo R., \href{https://doi.org/10.14311/AP.2013.53.0524}{``Search for the time dependence of radioactivity,''  \textit{Acta Polytechnica,}} 2013, {\bf 53}, 524-527. \textbf{Topics: B,E,NV,P}

\bibitem{Bellotti2015} Bellotti E., Broggini C., Di Carlo G, Laubenstein M., Menegazzo R., Pietroni M., \href{https://doi.org/10.1016/j.astropartphys.2014.05.006}{``Search for time modulations in the decay rate of \textsuperscript{40}K and \textsuperscript{232}Th,''  \textit{Astroparticle Physics,}} 2015, {\bf 61}, 82-87. \textbf{Topics: A,B,E,NV,P}

\bibitem{Bergeson} Bergeson S. D., Peatross J., Ware M. J., \href{https://doi.org/10.1016/j.physletb.2017.01.030}{``Precision long-term measurements of beta-decay-rate ratios in a controlled environment,''  \textit{Physics Letters B,}} 2017, {\bf 767}, 171-176. \textbf{Topics: B,E,NV}

\bibitem{Bikit} Bikit K., Nikolov J., Bikit I., Mrda D., Todorovic N., Forkapic S., Slivka J., Veskovic M., \href{https://doi.org/10.1016/j.astropartphys.2013.05.013}{``Reinvestigation of the irregularities in the \textsuperscript{3}H decay,''  \textit{Astroparticle Physics,}} 2013, {\bf 47}, 38-44. \textbf{Topics: B,E,V}

\item Boyarkin O. M., Boyarkina G. G., \href{https://doi.org/10.1016/j.astropartphys.2016.09.006}{``Influence of solar flares on behavior of solar neutrino flux,'' \textit{Astroparticle Physics,}} 2016, {\bf 85}, 39-42. \textbf{Topics: B,S,T,V}

\bibitem{Breur} Breur P. A., Nobelen J. C. P. Y, Baudis L., Brown A., Colijn A. P., Dressler R., Lang R. F., Massafferri A., Perci R., Pumar C., et al., \href{https://www.sciencedirect.com/science/article/abs/pii/S0927650520300049?via}{``Testing claims of the GW170817 binary neutron star inspiral affecting \textbeta -decay rates,'' \textit{Astroparticle Physics,}} 2020, {\bf 119}, 102431. \textbf{Topics: B,NV,P}

\bibitem{Buncher} Buncher J. B., \href{https://search-proquest-com.ezproxy.lib.purdue.edu/docview/858614008?accountid=13360}{``Phenomenology of Time-Varying Nuclear Decay Parameters,''} Ph.D. Thesis, Purdue University, 2010, unpublished. \textbf{Topics: P,S,V}

\item Cooper P. S., \href{https://doi.org/10.1016/j.astropartphys.2009.02.005}{``Searching for modifications to the exponential radioactive decay law with the Cassini spacecraft,'' \textit{Astroparticle Physics,}} 2009, {\bf 31}, Iss. 4, 267-269. \textbf{Topics: A,NV,T}

\item de Meijer R. J., Blaauw M., Smit F. D., \href{https://doi.org/10.1016/j.apradiso.2010.08.002}{``No evidence for antineutrinos significantly influencing exponential \textbeta \textsuperscript{+} decay,'' \textit{Applied Radiation and Isotopes,}} 2011, {\bf 69}, Iss. 2, 320-326. \textbf{Topics: A,NV,S,T}

\bibitem{Ellis} Ellis K. J., \href{https://iopscience.iop.org/article/10.1088/0031-9155/35/8/004}{``The effective half-life of a broad beam \textsuperscript{238}Pu/Be total body neutron irradiator,'' \textit{Physics in Medicine \& Biology,}} 1990, {\bf 35}, No. 8, 1079-1088. \textbf{Topics: E,V}

\item Elmaghraby E. K., \href{http://ptep-online.com/2017/PP-50-02.PDF}{``Configuration Mixing in Particle Decay and Reaction,'' \textit{Progress in Physics,}} 2017, {\bf 13}, Iss. 3, 150-155. \textbf{Topics: P,S,T,V}

\item Emery G. T., \href{https://www.annualreviews.org/doi/pdf/10.1146/annurev.ns.22.120172.001121}{``Perturbation of Nuclear Decay Rates,'' \textit{Annual Review of Nuclear Science,}} 1972, {\bf 22}, 165-202. \textbf{Topics: B,T,V}

\bibitem{Falkenberg} Falkenberg E. D., \href{https://www.semanticscholar.org/paper/Radioactive-Decay-Caused-by-Neutrinos-Falkenberg/3a21346203f836847e60016770d931b324a46be9}{``Radioactive Decay Caused by Neutrinos?,'' \textit{Apeiron,}} 2001, {\bf 8}, No. 2, 32-45. \textbf{Topics: B,E,EC,P,S,V}

\bibitem{GW} Fischbach E., Barnes V. E., Cinko N., Heim J., Kaplan H. B., Krause D. E., Leeman J. R., Mathews S. A., Mueterthies M. J., et al., \href{https://doi.org/10.1016/j.astropartphys.2018.06.001}{``Indications of an unexpected signal associated with the GW170817 binary neutron star inspiral,'' \textit{Astroparticle Physics,}} 2018, {\bf 103}, 1-6. \textbf{Topics: B,E,P,S,V}

\item Fischbach E., Buncher J. B., Gruenwald J. T., Jenkins J. H., Krause D. E., Mattes J. J., Newport J. R., \href{https://doi-org.ezproxy.lib.purdue.edu/10.1007/s11214-009-9518-5}{``Time-Dependent Nuclear Decay Parameters: New Evidence for New Forces?,'' \textit{Space Science Reviews,}} 2009, {\bf 145}, 285-335. \textbf{Topics: A,B,P,S,T,V}

\item Fischbach E., Krause D. E., Pattermann M., {``Comment on "Testing claims of the GW170817 binary neutron star inspiral affecting \textbeta-decay rates,'' \href{https://arxiv.org/pdf/2003.00092.pdf}{\underline{arXiv:2003.00092v1}} [nucl-ex], 2020. \textbf{Topics: P,S,V}

\item Fischbach E., Chen K. J., Gold R. E., Goldsten J. O., Lawrence D. J., McNutt Jr. R. J., Rhodes E. A., Jenkins J. H., Longuski J., \href{https://doi.org/10.1007/s10509-011-0808-5}{``Solar influence on nuclear decay rates: constraints from the MESSENGER mission,'' \textit{Astrophysics and Space Science,}} 2012, {\bf 337}, 39-45. \textbf{Topics: A,B,P,S,T,V}

\item Fischbach E., Jenkins J. H., Gruenwald J. T., Sturrock P. A., Javorsek II D.,  ``Evidence for Solar Influences on Nuclear Decay Rates,'' \href{https://doi.org/10.1142/9789814327688_0033}{In \textit{Proceedings of the fifth meeting on CPT and Lorentz Symmetry}}, (Edited by V. A. Kosteleck\'{y}), World Scientific, Singapore, 2010, 168-173. \textbf{Topics: A,B,P,S,T,V}

\item Fischbach E., Jenkins J. H., Sturrock P. A., ``Evidence for Time-varying Nuclear Decay Rates: Experimental Results and their Implications for New Physics,'' In \textit{Proceedings of the XLVIth Rencontres de Moriond, LaThuile, Italy,} (Edited by Etienne Augé, Jacques Dumarchez, Jean Trân Thanh Vân), \href{https://arxiv.org/abs/1106.1470}{\underline{arXiv:1106.1470}} [nucl-ex], 2011. \textbf{Topics: P,S,T,V}

\bibitem{Goddard} Goddard B., Hitt G. W., Solodov A. A., Bridi D., Isakovic A. F., El-Khazali R., Abulail A., \href{https://doi.org/10.1016/j.nima.2015.12.026}{``Experimental setup and commissioning baseline study in search of time-variations in beta-decay half-lives,'' \textit{Nuclear Instruments and Methods in Physics Research A,}} 2016, {\bf 812}, 60-67. \textbf{Topics: E,NV}

\bibitem{Goddard_Au198}Goddard B., Golovko V. V., Iacab V. E., Hardy J. C., \href{https://doi-org.ezproxy.lib.purdue.edu/10.1140/epja/i2007-10509-0}{``The half-life of \textsuperscript{198}Au: High-precision measurement shows no temperature dependence,'' \textit{European Physical Journal A,}} 2007, {\bf 34}, 271-274. \textbf{Topics: E,NV}

\item Greenland P. T., \href{https://doi-org.ezproxy.lib.purdue.edu/10.1038/335298a0}{``Seeking non-exponential decay,'' \textit{Nature,}} 1988, {\bf 335}, 298. \textbf{Topics: T,NV}

\item Hahn H. P., Born H. J., Kim J. I., \href{https://doi.org/10.1524/ract.1976.23.1.23}{``Survey on the Rate Perturbation of Nuclear Decay'' \textit{Radiochimica Acta,}} 1976, {\bf 23}, 23-37. \textbf{Topics: NV,R}

\item Hardy J. C, Goodwin J. R., Golovko V. V., Iacob V. E., \href{https://doi.org/10.1016/j.apradiso.2009.11.047}{``Tests of nuclear half-lives as a function of the host medium and temperature: Refutation or recent claims'' \textit{Applied Radiation and Isotopes,}} 2010, {\bf 68}, 1550-1554. \textbf{Topics: B,E,NV,P,S}

\bibitem{Hardy} Hardy J. C, Goodwin J. R., Iacob V. E., \href{https://doi.org/10.1016/j.apradiso.2012.02.021}{``Do radioactive half-lives vary with the Earth-to-Sun distance?'' \textit{Applied Radiation and Isotopes,}} 2012, {\bf 70}, Iss. 9, 1931-1933. \textbf{Topics: B,NV,P,S}

\bibitem {Heim} Heim J. M., \href{https://search-proquest-com.ezproxy.lib.purdue.edu/docview/1781664360?accountid=13360}{``The Determination of the Half-life of Si-32 and Time Varying Nuclear Decay,''}  Ph.D. Thesis, Purdue University, 2015, unpublished. \textbf{Topics: B,S,V}

\item Jaubert F., Tartès I., Cassette P., \href{https://doi.org/10.1016/j.apradiso.2006.02.081}{``Quality control of liquid scintillation counters,'' \textit{Applied Radiation and Isotopes,}} 2006, {\bf 64}, Iss. 10, 1163-1170. \textbf{Topics: }

\item Javorsek II D., Brewer M. C., Buncher J. B., Fischbach E., Gruenwald J. T., Heim J., Hoft A. W., Horan T. J., Kerford J. L., Kohler M., et al., \href{https://doi.org/10.1007/s10509-012-1148-9}{``Study of nuclear decays during a solar eclipse: Thule Greenland 2008,'' \textit{Astrophysics and Space Science,}} 2012, {\bf 342}, 9-13. \textbf{Topics: P,S,T,V}

\bibitem{Javorsek} Javorsek D., Kerford J. L., Stewart C. A., Buncher J. B., Fischbach E., Gruenwald J. T., Heim J., Hoft A. W., Horan T. J., Jenkins J. H., et al., \href{https://doi.org/10.1063/1.3480162}{``Preliminary Results from Nuclear Decay Experiments Performed During the Solar Eclipse of August 1, 2008,'' \textit{AIP Conference Proceedings,}} 2010, {\bf 1265}, Iss. 1, 178-179. \textbf{Topics: A,B,P,S,V}

\item Javorsek II D., Sturrock P. A., Buncher J. B., Fischbach E., Gruenwald J. T., Hoft A. W., Horan T. J., Jenkins, J. H., et al., \href{https://doi.org/10.1063/1.3293804}{``Investigation of Periodic Nuclear Decay Data with Spectral Analysis Techniques,'' \textit{AIP Conference Proceedings,}} 2009, {\bf 1182}, Iss 292. \textbf{Topics: }

\item Javorsek II D., Sturrock P. A., Lasenby R. N., Lasenby A. N., Buncher J. B., Fischbach E., Gruenwald J. T., Hoft A. W., Horan T. J., Jenkins, J. H., et al., \href{https://doi.org/10.1016/j.astropartphys.2010.06.011}{``Power spectrum analyses of nuclear decay rates,'' \textit{Astroparticle Physics,}} 2010, {\bf 34}, Iss. 3, 173-178. \textbf{Topics: A,B,P,V}

\item Javorsek II D., Sturrock P. A., Lasenby R. N., Lasenby A. N., Buncher J. B., Fischbach E., Gruenwald J. T., Jenkins, J. H., Lee R. H., Mattes J. J., et al., \href{https://doi.org/10.1063/1.3480152}{``Periodicities in Nuclear Decay Data: Systematic Effects or New Physics?,'' \textit{AIP Conference Proceedings,}} 2010, {\bf 1265}, Iss. 1, 144-147. \textbf{Topics: A,B,P,S,V}

\bibitem{Jenkins_solar_flare} Jenkins, J. H., Fischbach E., \href{https://doi.org/10.1016/j.astropartphys.2009.04.005} {``Perturbation of nuclear decay rates during the solar flare of 2006 December 13,'' \textit{Astroparticle Phyiscs,}} 2009, {\bf 31}, Iss. 6, 407-411. \textbf{Topics: B,E,P,S,V}

\bibitem{Jenkins_Solar} Jenkins, J. H., Fischbach E., Buncher J. B., Gruenwald J. T., Krause D. E., Mattes J. J., \href{https://doi.org/10.1016/j.astropartphys.2009.05.004}{``Evidence of correlations between nuclear decay rates and Earth-Sun distance,'' \textit{Astroparticle Physics,}} 2009, {\bf 32}, No. 1, 42-46. \textbf{Topics: A,B,P,V}

\bibitem{Jenkins_Cs137} Jenkins, J. H., Fischbach E., Javorsek II D., Lee R. H., Sturrock P. A., \href{https://doi.org/10.1016/j.apradiso.2012.12.010} {``Concerning the time dependence of the decay rate of \textsuperscript{137}Cs,'' \textit{Applied Radiation and Isotopes,}} 2013, {\bf 74}, 50-55. \textbf{Topics: A,B,P,V}

\bibitem{Jenkins_arXiv} Jenkins, J. H., Fischbach E., Sturrock P. A., Mundy D. W., ``Analysis of Experiments Exhibiting Time-Varying Nuclear Decay Rates: Systematic Effects or New Physics?,''  \href{https://arxiv.org/abs/1106.1678} {\underline{arXiv:1106.1678}} [nucl-ex], 2011. \textbf{Topics: A,B,P,S,V}

\bibitem{Jenkins_Astroparticle} Jenkins, J. H., Herminghuysen K. R., Blue T. E., Fischbach E., Javorsek II D., Kauffman A. C., Mundy D. W., Sturrock P. A., Talnagi J. W., \href{https://doi.org/10.1016/j.astropartphys.2012.07.008} {``Additional experimental evidence for a solar influence on nuclear decay rates,'' \textit{Astroparticle Physics,}} 2012, {\bf 37}, 81-88. \textbf{Topics: B,P,S,V}

\bibitem{JenkinsSi32} Jenkins J. H., Mundy D. W., Fischbach E., \href{https://doi.org/10.1016/j.nima.2010.03.129}{``Analysis of environmental influences in nuclear half-life measurements exhibiting time-dependent decay rates,'' \textit{Nuclear Instruments and Methods in Physics Research Section A: Accelerators, Spectrometers, Detectors and Associated Equipment,}} 2010, {\bf 620}, Iss. 2-3, 332-342. \textbf{Topics: A,B,P,V}

\item Jerome S., Bobin C., Cassette P., Dersch R., Galea R., Liu H., Honig A., Keightley J., Kossert K., Liang J., et al., \href{https://doi.org/10.1016/j.apradiso.2019.108837}{``Half-life determination and comparison of activity standards of \textsuperscript{231}Pa,'' \textit{Applied Radiation and Isotopes,}} 2020, {\bf155}, 108837.  \textbf{Topics: NV,P}

\bibitem{Kay} Kay M. J., \href{https://doi.org/10.25394/PGS.7403810.v1}{``New Methodologies for measuring and monitoring nuclear decay parameters for time dependent behavior,'' } Ph.D. Thesis, Purdue University, 2018, unpublished. \textbf{Topics: B,E,P,S,V}

\item Khavroshkin O. B., Tsyplakov V. V., \href{https://doi.org/10.4236/oalib.1104869}{``Five Years after Discovery Abnormal Neutrino Radioactive-Isotope (ANRI) Absorption,'' \textit{Open Access Library Journal,}} 2018, {\bf 5}, e4869. \textbf{Topics: B,E,P,S,V}

\bibitem{Knezevic} Knezevic J., Mrdja D., Bikit-Schroeder K., Hansman J., Bikit I., Slivka J., \href{https://doi.org/10.1016/j.apradiso.2020.109178}{``Search for variances of \textsuperscript{22}Na decay constant,'' \textit{Applied Radiation and Isotopes,}} 2020, {\bf 163}, 109178. \textbf{Topics: B,NV,P}

\bibitem{KossertFe} Kossert K., \href{https://doi.org/10.1016/j.apradiso.2019.108931}{``TDCR measurements to determine the half-life of \textsuperscript{55}Fe,'' \textit{Applied Radiation and Isotopes,}} 2020, {\bf 155}, 108931. \textbf{Topics: E,EC,NV}

\bibitem{Kossert_Sr} Kossert K., Nähle O. J.,  \href{https://doi.org/10.1016/j.astropartphys.2015.03.003}{``Disproof of solar influence on the decay rates of \textsuperscript{90}Sr/\textsuperscript{90}Y,'' \textit{Astroparticle Physics,}} 2015, {\bf 69}, 18-23. \textbf{Topics: B,E,NV,P,S}

\bibitem{KossertCl36} Kossert K., Nähle O. J.,  \href{https://doi.org/10.1016/j.astropartphys.2014.02.001}{``Long-term measurements of \textsuperscript{36}Cl to investigate potential solar influence on the decay rate,'' \textit{Astroparticle Physics,}} 2014, {\bf 55}, 33-36. \textbf{Topics: B,NV,P,S}

\bibitem{Krause} Krause D. E., Rogers B. A., Fischbach E., Buncher J. B., Ging A., Jenkins J. H., Longuski J. M., Strange N., Sturrock P. A., \href{https://doi.org/10.1016/j.astropartphys.2012.05.002}{``Searches for solar-influenced radioactive decay anomalies using spacecraft RTGs,'' \textit{Astroparticle Physics,}} 2012, {\bf 36}, Iss. 1, 51-56. \textbf{Topics: P,A,S,V}

\bibitem{LeePaper} Lee R. H., Fischbach E., Gruenwald J. T., Javorsek II D., Sturrock P. A., \href{https://journals.ke-i.org/qpr/article/download/1258/925}{``Spectral Content in \textsuperscript{3}H and \textsuperscript{14}C Decays: A review of Five Experiments,'' \textit{Quarterly Physics Review}} 2017, {\bf 3}, Iss. 2. \textbf{Topics: P,R,V}

\bibitem{Lee} Lee R. H., Javorsek II D., Morris D., \href{https://www.ctbto.org/fileadmin/user_upload/SnT2013/Slides/Wednesday/T3-O28_Lee-Javorsek.pdf}{``Stability of the IMS Radionuclide Detector Network and Lessons Learnt for Exotic Physics Searches,'' \textit{Comprehensive Nuclear-Test-Ban Treaty: Science and Technology Conference,}} 2013. \textbf{Topics: P,R,V}

\item Lindstrom R. M., \href{https://doi-org.ezproxy.lib.purdue.edu/10.1007/s10967-016-4912-4}{``Believable statement of uncertainty and believable science,'' \textit{Journal of Radioanalytical and Nuclear Chemistry,}} 2017, {\bf 311}, 1019-1022. \textbf{Topics: B,P,S,T,V}

\bibitem{LindstromAu} Lindstrom R. M., Fischbach E., Buncher J. B., Greene G. L., Jenkins J. H., Krause D. E., Mattes J. J., Yue A., \href{https://doi.org/10.1016/j.nima.2010.06.270}{``Study of the dependence of \textsuperscript{198}Au half-life on source geometry,'' \textit{Nuclear Instruments and Methods in Physics Research Section A,}} 2010, {\bf 622}, Iss. 1, 93-96. \textbf{Topics: B,P,S,T,V}

\bibitem{Lindstrom} Lindstrom R. M., Fischbach E., Buncher J. B., Jenkins J. H., Yue A., \href{https://doi.org/10.1016/j.nima.2011.08.046}{``Absence of a self-induced decay effect in \textsuperscript{198}Au,'' \textit{Nuclear Instruments and Methods in Physics Research Section A,}} 2011, {\bf 659}, Iss. 1, 269-271. \textbf{Topics: B,P,S,V}

\bibitem{Lobashev} Lobashev V. M., Aseev V. N., Belesev A. I., Berlev A. I., Geraskin E. V., Golubev A. A., Kazachenko O. V., Kuznetsov Y. E., Ostroumov R. P., Rivkis L. A., et al., \href{https://doi.org/10.1016/S0370-2693(99)00781-9}{``Direct search for mass of neutrino and anomaly in the tritium beta-spectum,'' \textit{Physics Letters B,}} 1999, {\bf 460}, Iss. 1-2, 227-235. \textbf{Topics: B,E,V}

\item Lobashev V. M., Aseev V. N., Belesev A. I., Berlev A. I., Geraskin E. V., Golubev A. A., Kazachenko O. V., Kuznetsov Y. E., Ostroumov R. P., Rivkis L. A., et al., \href{https://doi.org/10.1016/S0920-5632(00)00952-X}{``Direct Search for Neutrino Mass and Anomaly in the Tritium Beta-Spectrum: Status of ''Troitsk Neutrino Mass'' Experiment,'' \textit{Nuclear Physics B,}} 2001, {\bf 91}, 280-286. \textbf{Topics: }

\item Mattes J. J., \href{https://search.proquest.com/docview/1435651033?accountid=13360}{``Detecting Relic Neutrinos Through Coherent Processes,''}  Ph.D. Thesis, Purdue University, 2013, unpublished. \textbf{Topics: S,V}

\item Mayburov S., \href{https://doi.org/10.1007/s10773-019-04237-x}{``Nuclear Decay Parameter Oscillations as Possible Signal of Quantum-Mechanical Nonlinearity,'' \textit{International Journal of Theoretical Physics,}} 2019. \textbf{Topics: A,B,T,V}

\bibitem{McKnight} McKnight Q., Bergeson S. D., Peatross J., Ware M. J., \href{https://doi.org/10.1016/j.apradiso.2018.09.021}{``2.7 years of beta-decay-rate ratio measurements in a controlled environment,'' \textit{Applied Radiation and Isotopes,}} 2018, {\bf 142}, 113-119. \textbf{Topics: NV,T}

\item Meier M. M. M., Wieler R., \href{https://doi.org/10.1016/j.astropartphys.2014.01.004}{``No evidence for a decrease of nuclear decay rates with increasing heliocentric distance based on radiochronology of meteorites,'' \textit{Astroparticle Physics,}} 2014, {\bf 55}, 63-75. \textbf{Topics: A,B,NV,P}

\bibitem{Sanchez} Milián-Sánchez V., Scholkmann F., Fernández de Córdoba P., Mocholí-Salcedo A., Mocholí F., Iglesias-Martínez M. E., Castro-Palacio J. C., Kolombet V. A., Panchelyuga V. A., Verdú G., \href{https://doi.org/10.1038/s41598-020-64497-0}{``Fluctuations in measured radioactive decay rates inside a modified Faraday cage: Correlations with space weather,'' \textit{Scientific Reports,}} 2020, {\bf 10}, 8525. \textbf{Topics: E,V}

\item Mohsinally T., Fancher S., Czerny C., Fischbach E., Gruenwald T., Heim J., Jenkins J. H., Nistor J., O'Keefe D., \href{https://doi.org/10.1016/j.astropartphys.2015.10.007}{``Evidence for correlations between fluctuations in \textsuperscript{54}Mn decay rates and solar storms,'' \textit{Astroparticle Physics,}} 2016, {\bf 75}, 29-37. \textbf{Topics: B,EC,S,T,V}

\bibitem{MohsinallyThesis} Mohsinally T., \href{https://docs.lib.purdue.edu/open_access_dissertations/521/}{``An investigation into the phenomenological relation between solar activity and nuclear beta-decay rates,''} Ph.D. Thesis, Purdue University, 2015, unpublished. \textbf{Topics: B,P,S,V}

\item Mueterthies  M. J.,  Krause D.E., Longman A.,  Barnes V.E,  Fischbach E., ``Is there a signal for Lorentz non-invariance in existing radioactive decay data?'' In \textit{Proceedings of the Seventh meeting on CPT and Lorentz symmetry}, (Edited by V. A. Kosteleck\'{y}), World Scientific, Singapore, 2017, 197-200. \href{https://arxiv.org/abs/1407.4144}{\underline{arXiv:1607.03541}} [nep-ph], 2016. \textbf{Topics: P,S,V}

\bibitem{Nahle} Nähle O., Kossert K., \href{https://doi.org/10.1016/j.astropartphys.2014.11.005}{``Comment on 'Comparative study of beta-decay data for eight nuclides measured at the Physikalisch-Technische Bundesanstalt,''' \textit{Astroparticle Physics,}} 2015, {\bf 66}, 8-10. \textbf{Topics: B,NV,P}

\item Nistor J., \href{https://docs.lib.purdue.edu/open_access_dissertations/1314/}{``Direct and Indirect Searches for Anomalous Beta Decay,''} Ph.D. Thesis, Purdue University, 2015, unpublished. \textbf{Topics: B,S,V}

\item Nistor J., Heim J., Fischbach E., Jenkins J. H., Sturrock P. A., ``Phenomenology of Rate–Related Nonlinear Effects in Nuclear Spectroscopy,'' \href{https://arxiv.org/abs/1407.4144}{\underline{arXiv:1407.4144}} [nucl-ex], 2014. \textbf{Topics: P,S,V}

\item Nistor J., Fischbach E., Gruenwald J. T., Javorsek D., Jenkins J. H., Lee R. H., ``Time-Varying nuclear decay parameters and Dark Matter,'' In \textit{Proceedings on the Sixth Meeting on CPT and Lorentz Symmetry,}} (Edited by V. A. Kosteleck\'{y}), World Scientific, 2014, 208-211. \href{https://arxiv.org/pdf/1307.7620.pdf}{\underline{arXiv:1307.7620}} [hep-ph] \textbf{Topics: P,S,V}

\bibitem{Norman} Norman E. B., Browne E., Shugart H. A., Joshi T. H., Firestone R. B., \href{https://doi.org/10.1016/j.astropartphys.2008.12.004}{``Evidence against correlations between nuclear decay rates and Earth–Sun distance,''' \textit{Astroparticle Physics,}} 2009, {\bf 31}, Iss. 2, 135-137. \textbf{Topics: A,B,NV,P}

\item Norman E. B., Gazes S. B., Crane S. G., Bennett D. A., \href{https://doi.org/10.1103/PhysRevLett.60.2246}{``Tests of the Exponential Decay Law at Short and Long Times,''' \textit{Physical Review,}} 1988, {\bf 60}, Iss. 22, 2246-2249. \textbf{Topics: NV,P,T}

\item Norman E. B., Sur B., Lesko K. T., Larimer R. M., DePaolo D. J., Owens T. L., \href{https://doi.org/10.1016/0370-2693(95)00818-6}{``An improved test of the exponential decay law,'' \textit{Physics Letters B,}} 1995, {\bf 357}, Iss. 4, 521-525. \textbf{Topics: NV,P,T}

\bibitem{O'Keefe} O'Keefe D., Morreale B. L., Lee R. H., Buncher J. B., Jenkins J. H., Fischbach E., Gruenwald T., Javorsek D., Sturrock P. A.,  \href{https://doi.org/10.1007/s10509-012-1336-7}{``Spectral content of \textsuperscript{22}Na/\textsuperscript{44}Ti decay data: Implications for a solar influence,'' \textit{Astrophysics and Space Science,}} 2013, {\bf 344}, 297-303. \textbf{Topics: P,S,V}

\item Papaloizou J., Pringle J. E.,  \href{https://doi.org/10.1093/mnras/182.3.423}{``Non-radial oscillations of rotating stars and their relevance to the short-period oscillations of cataclysmic variables,'' \textit{Monthly Notices of the Royal Astronomical Society,}} 1978, {\bf 182}, Iss. 3, 423-442. \textbf{Topics: T,V}

\item Parkhomov A. G., \href{http://ripublication.com/ijpap/1001.pdf}{``Bursts of Count Rate of Beta-Radioactive Sources during Long-Term Measurements,'' \textit{International Journal of Pure and Applied Physics,}} 2005, {\bf 1}, No. 2, 119-128. \textbf{Topics: B,E,P,S,V}

\item Parkhomov A. G., \href{https://www.scirp.org/journal/PaperInformation.aspx?PaperID=8635}{``Deviations from Beta Radioactivity Exponential Drop,'' \textit{Journal of Modern Physics,}} 2011, {\bf 2}, No. 11, 1310-1317. \textbf{Topics: A,B,P,V}

\item Parkhomov A. G., ``Effect of radioactivity decrease. Is there a link with solar flares?,'' \href{https://arxiv.org/abs/1006.2295}{\underline{arXiv:1006.2295}} [physics.gen-ph], 2010. \textbf{Topics: B,P,S,V}

\item Parkhomov A. G., ``Influence of Relic Neutrinos on Beta Radioactivity,'' \href{https://arxiv.org/abs/1010.1591}{\underline{arXiv:1010.1591}} [physics.gen-ph], 2010. \textbf{Topics: B,P,S,V}

\bibitem{Parkhomov_arXiv} Parkhomov A. G., ``Periods Detected During Analysis of Radioactivity Measurements Data,'' \href{https://arxiv.org/abs/1012.4174}{\underline{arXiv:1012.4174}} [physics.gen-ph], 2010. \textbf{Topics: B,P,V}

\bibitem{Parkhomov_alpha_beta} Parkhomov A. G., ``Researches of alpha and beta radioactivity at long-term observations,'' \href{https://arxiv.org/abs/1004.1761}{\underline{arXiv:1004.1761}} [physics.gen-ph], 2010. \textbf{Topics: A,B,P,V}

\item Picolo J. L., \href{https://doi.org/10.1016/S0168-9002(96)80029-5}{``Absolute measurement of radon 222 activity,'' \textit{Nuclear Instruments and Methods in Physics Research Section A: Accelerators, Spectrometers, Detectors and Associated Equipment,}} 1996, {\bf 369}, Iss. 2-3, 452-457. \textbf{Topics: B,P,V}

\item Pommé S., \href{https://doi.org/10.1016/j.nima.2020.163933}{``Comparing significance criteria for cyclic modulations in time series,'' \textit{Nuclear Instruments and Methods in Physics Research Section A: Accelerators, Spectrometers, Detectors and Associated Equipment,}} 2020, {\bf 968}, 163933. \textbf{Topics: NV,T}

\item Pommé S., \href{https://iopscience.iop.org/article/10.1088/0026-1394/44/4/S03/pdf?casa_token=SK1WFJBqVN0AAAAA:nSGYkM9CHpV7i43ipUIz7fwq5T_NRY4uKjo2hMYqgxmOcWP4GPHtPNDypDacCp01n6W9Qjz6XBN7MjWvvnnv}{``Methods for primary standardization of activity,'' \textit{Metrologia,}} 2007, {\bf 44}, No. 4, 17-26. \textbf{Topics: A,B,NV,P,T}

\item Pommé S., \href{https://pubs.acs.org/doi/abs/10.1021/bk-2007-0945.ch020}{``Problems with the Uncertainty Budget of Half-life Measurements,'' \textit{Applied Modeling and Computations in Nuclear Science,}} 2006, {\bf 945}, Ch 20, 282-292. \textbf{Topics: NV,P}

\item Pommé S., \href{https://doi.org/10.1140/epjc/s10052-019-6597-7}{``Solar influence on radon decay rates: irradiance or neutrinos?,'' \textit{European Physical Journal C,}} 2019, {\bf 79}, No. 73. \textbf{Topics: A,B,G,NV,P,S,T}

\item Pommé S., \href{https://iopscience.iop.org/article/10.1088/0026-1394/52/3/S51/meta}{``The uncertainty of the half-life,'' \textit{Metrologia,}} 2015, {\bf 52}, 51 - 65. \textbf{Topics: NV,P,R,T}

\item Pommé S., Camps J., Van Ammel R., Paepen J., \href{https://akjournals.com/view/journals/10967/276/2/article-p335.xml}{``Protocol for uncertainty assessment of half-lives,'' \textit{Journal of Radioanalytical and Nuclear Chemistry,}} 2008, {\bf 276}, Iss. 2, 335-339. \textbf{Topics: NV,P,T}

\item Pommé S., De Hauwere T.,\href{https://doi.org/10.1016/j.apradiso.2020.109046}{``Derivation of an uncertainty propagation factor for half-life determinations,'' \textit{Applied Radiation and Isotopes,}} 2020, {\bf 158}. \textbf{Topics: NV,T}

\item Pommé S., De Hauwere T., \href{https://www.sciencedirect.com/science/article/pii/S0168900219316171}{``On the significance of modulations in time series,'' \textit{Nuclear Instruments and Methods in Physics Research Section A: Accelerators, Spectrometers, Detectors and Associated Equipment,}} 2020, {\bf 956}, 163377. \textbf{Topics: NV,T}

\item Pommé S., Kossert K., Nähle O., \href{https://doi.org/10.1007/s11207-017-1187-z}{``On the Claim of Modulations in \textsuperscript{36}Cl Beta Decay and Their Association with Solar Rotation,'' \textit{Solar Physics,}} 2017, {\bf 292}, No. 162. \textbf{Topics: NV,P,S}

\item Pommé S., Lutter G., Marouli M., Kossert K., Nähle O., \href{https://doi.org/10.1016/j.astropartphys.2018.11.005}{``A reply to the rebuttal by Sturrock et al.,'' \textit{Astroparticle Physics,}} 2019, {\bf 107}, 22-25. \textbf{Topics: B,NV,P}

\item Pommé S., Lutter G., Marouli M., Kossert K., Nähle O., \href{https://doi.org/10.1016/j.astropartphys.2017.10.011}{``On the claim of modulations in radon decay and their association with solar rotation,'' \textit{Astroparticle Physics,}} 2018, {\bf 97}, 38-45. \textbf{Topics: B,NV,P,S}

\item Pommé S., Marouli M., Suliman G., Dikmen H., Van Ammel R., Jobbágy V., Dirican A., Stroh H., Paepen J., Bruchertseifer F., et al., \href{https://doi.org/10.1016/j.apradiso.2012.07.014}{``Measurement of the \textsuperscript{225}Ac half-life,'' \textit{Applied Radiation and Isotopes,}} 2012, {\bf 70},  Iss. 11, 2608-2614. \textbf{Topics: T,NV}

\item Pommé S., Stroh H., Altzitzoglou T., Paepen J., Van Ammel R., Kossert K., Nähle O., Keightley J. D., Ferreira K. M., Verheyen L., et al. \href{https://doi.org/10.1016/j.apradiso.2017.09.002} {``Is decay constant?,'' \textit{Applied Radiation and Isotopes,}} 2018, {\bf 134}, 6-12. \textbf{Topics: A,B,EC,NV,P}

\item Pommé S., Stroh H., Paepen J., Van Ammel R., Marouli M., Altzitzoglou T., Hult M., Kossert K., Nähle O., Schrader H., et al. \href{https://doi.org/10.1016/j.physletb.2016.08.038}{``Evidence against solar influence on nuclear decay constants,'' \textit{Physics Letters B,}} 2016, {\bf 761}, 281-286. \textbf{Topics: A,B,NV,P,R,S}

\item Pommé S., Stroh H., Paepen J., Van Ammel R., Marouli M., Altzitzoglou T., Hult M., Kossert K., Nähle O., Schrader H., et al. \href{https://iopscience.iop.org/article/10.1088/1681-7575/54/1/1/meta}{``On decay constants and orbital distance to the Sun- part I: alpha decay,'' \textit{Metrologia,}} 2017, {\bf 54}, No. 1, 1-18. \textbf{Topics: A,NV,P}

\item Pommé S., Stroh H., Paepen J., Van Ammel R., Marouli M., Altzitzoglou T., Hult M., Kossert K., Nähle O., Schrader H., et al. \href{https://iopscience.iop.org/article/10.1088/1681-7575/54/1/19/meta}{``On decay constants and orbital distance to the Sun- part II: beta minus decay,'' \textit{Metrologia,}} 2017, {\bf 54}, No. 1, 19-35. \textbf{Topics: B,NV,P}

\item Pommé S., Stroh H., Van Ammel R., \href{https://doi.org/10.1016/j.apradiso.2019.01.008}{``The \textsuperscript{55}Fe half-life measured with a pressurised proportional counter,'' \textit{Applied Radiation and Isotopes,}} 2019, {\bf 148}, 27-34. \textbf{Topics: E,EC,NV}

\item Pommé S., Stroh H., Van Ammel R., \href{https://iopscience.iop.org/article/10.1088/1681-7575/54/1/36/meta}{``On decay constants and orbital distance to the Sun- part III: beta plus and electron capture decay,'' \textit{Metrologia,}} 2017, {\bf 54}, No. 1, 36-50. \textbf{Topics: B,EC,NV,P}

\item Pons D. J., Pons A. D., Pons A. J., \href{http://www.ccsenet.org/journal/index.php/apr/article/view/46281}{``Hidden Variable Theory Supports Variability in Decay Rates of Nuclides,'' \textit{Applied Physics Research}} 2015, {\bf 7}, No. 7. \textbf{Topics: P,V}

\item Sanders A. J., ``Implications for \textsuperscript{14}C Dating of the Jenkins-Fischbach Effect and Possible Fluctuation of the Solar Fusion Rate,'' \href{https://arxiv.org/abs/0808.3986}{\underline{arXiv:0808.3986}} [astro-ph], 2011. \textbf{Topics: P,S,V}

\bibitem{Schrader} Schrader H., \href{https://doi.org/10.1016/j.apradiso.2009.11.033}{``Half-life measurements of long-lived radionuclides—New data analysis and systematic effects,'' \textit{Applied Radiation and Isotopes,}} 2010, {\bf 68}, Iss. 7-8, 1583-1590. \textbf{Topics: NV,P}

\bibitem{Schrader_F18} Schrader H., \href{https://doi.org/10.1016/j.apradiso.2003.11.039}{``Half-life measurements with ionization chambers—A study of systematic effects and results,'' \textit{Applied Radiation and Isotopes,}} 2004, {\bf 60}, Iss. 2-4, 317-323. \textbf{Topics: E,NV}

\bibitem{Schrader_Eu} Schrader H., \href{https://doi.org/10.1088/0026-1394/44/4/S07}{``Ionization chambers,'' \textit{Metrologia,}} 2007, {\bf 44}, No. 4, 53-66. \textbf{Topics: E,NV}

\bibitem{Schrader_Ra} Schrader H., \href{https://doi.org/10.1016/0168-9002(92)90122-K}{``Measurement of small currents using a modified Townsend method with fast, programmable A/D and D/A converters,'' \textit{Nuclear Instruments and Methods in Physics Research Section A: Accelerators, Spectrometers, Detectors and Associated Equipment,}} 1992, {\bf 312}, Iss. 1-2, 34-38. \textbf{Topics: E,NV}

\bibitem{Schrader_2016} Schrader H., \href{https://doi.org/10.1016/j.apradiso.2016.05.001}{``Seasonal variations of decay rate measurement data and their interpretation,'' \textit{Applied Radiation and Isotopes,}} 2016, {\bf 114}, 202-213. \textbf{Topics: E,NV}

\bibitem{Siegert} Siegert H., Schrader H., Sch\"{o}tzig U., \href{https://doi.org/10.1016/S0969-8043(97)10082-3}{``Half-life measurements of Europium radionuclides and the long-term stability of detectors,'' \textit{Applied Radiation and Isotopes,}} 1998, {\bf 49}, Iss. 9-11, 1397-1401. \textbf{Topics: P,B,EC,V}

\item Singleton D., Inan N., Chiao R. Y., \href{https://doi.org/10.1016/j.physleta.2015.01.028}{``Neutrino induced decoherence and variation in nuclear decay rates,'' \textit{Physics Letters A,}} 2015, {\bf 379}, Iss. 12-13, 941-946. \textbf{Topics: T,V}

\bibitem{Semkow} Semkow T. M., Haines D. K., Beach S. E., Kilpatrick B. J., Khan A. J., O'Brien K., \href{https://doi.org/10.1016/j.physletb.2009.04.051}{``Oscillations in radioactive exponential decay,'' \textit{Physics Letters B,}} 2009, {\bf 675}, Iss. 5, 415-419. \textbf{Topics: B,E,NV}

\item Shirai T., \href{https://doi.org/10.1023/B:SOLA.0000043565.83411.ec}{``Time variation of the solar neutrino fluxes from Super-Kamiokande data,'' \textit{Solar Physics,}} 2004, {\bf 222}, 199-201. \textbf{Topics: P,V,S}

\bibitem{Shnoll} Shnoll S. E., Kolombet V. A., Pozharskii E. V., Zenchenko T. A., Zvereva I. M., Konradov A., \href{https://iopscience.iop.org/article/10.1070/PU1998v041n10ABEH000463/meta}{``Realization of discrete states during fluctuations in macroscopic processes,'' \textit{Uspekhi Fizicheskikh Nauk and Russian Academy of Science,}} 1998, {\bf 41}, No. 10, 1025-1035. \textbf{Topics: E,V}

\item Stancil D. D., Yegen S. B., Dickey D. A., Gould C. R., \href{https://doi.org/10.1016/j.rinp.2016.12.051}{``Search for possible solar influences in Ra-226 decays,'' \textit{Results in Physics,}} 2017, {\bf 7}, 385-406. \textbf{Topics: P,S,V}

\item Steinitz G., Flore N., Piatibratova O., \href{https://royalsocietypublishing.org/doi/10.1098/rspa.2018.0240}{``Indications for solar influence on radon and thoron in the atmosphere, Arad, Romania,'' \textit{Proceedings of the Royal Society A,}} 2018, {\bf 474}. \textbf{Topics: S,T,V}

\bibitem{Steinitz_Ra226} Steinitz G.,Kotlarsky P., Piatibratova O., \href{https://doi.org/10.1140/epjst/e2015-02403-2}{``Observations of the relationship between directionality and decay rate of radon in a confined experiment,'' \textit{European Physical Journal Special Topics,}} 2015, {\bf 224}, 731-740. \textbf{Topics: E,P,S,V}

\item Steinitz G., Martin-Luis M. C., Piatibratova O., \href{https://doi.org/10.1140/epjst/e2015-02399-5}{``Indications for solar influence on radon signal in the subsurface of Tenerife (Canary Islands, Spain),'' \textit{European Physical Journal Special Topics,}} 2015, {\bf 224}, 687-695. \textbf{Topics: P,S,V}

\item Steinitz G., Piatibratova O., Gazit-Yaari N., \href{https://doi.org/10.1098/rspa.2013.0411}{``Influence of a component of solar irradiance on radon signals at 1 km depth, Gran Sasso, Italy,'' \textit{Royal Society Publishing A,}} 2013, {\bf 469}. \textbf{Topics: A,E,P,S,V}

\item Steinitz G., Piatibratova O., Malik U., \href{https://doi.org/10.1140/epjst/e2015-02396-8}{``Observations on the spatio-temporal patterns of radon along the western fault of the Dead Sea Transform, NW Dead Sea,'' \textit{European Physical Journal Special Topics,}} 2015, {\bf 224}, 629-639. \textbf{Topics: A,E,P,S,V}

\bibitem{Steinitz} Steinitz G., Piatibratova O., Kotlarsky P., \href{https://doi.org/10.1016/j.jenvrad.2011.04.002}{``Possible effect of solar tides on radon signals,'' \textit{Journal of Environmental Radioactivity,}} 2011, {\bf 102}, Iss. 8, 749-765. \textbf{Topics: A,B,E,P,V}

\bibitem{Steinitz_ESS} Steinitz G., Sturrock P., Fischbach E., Piatibratova O., \href{https://doi.org/10.1002/essoar.a0e6de6afdf78d90.905b86c97fa74b0c.1}{``Indications for non-terrestrial influences on radon signals from a multi-year enhanced confined experiment,'' \textit{Earth and Space Science Open Access,}} 2018. \textbf{Topics: P,S,V}

\item Sturrock P. A., Bertello L., \href{https://iopscience.iop.org/article/10.1088/0004-637X/725/1/492/pdf}{''Power spectrum analysis of Mount Wilson solar diameter measurements: Evidence for solar internal r-mode oscillations,'' \textit{Astrophysical Journal,}} 2010, {\bf 725}, 492-495. \textbf{Topics: S,T,V}

\item Sturrock P. A., Bertello L., Fischbach E., Javorsek II D., Jenkins J. H., Kosovichev A., Parkhomov A. G., \href{https://doi.org/10.1016/j.astropartphys.2012.11.011}{``An analysis of apparent r-mode oscillations in solar activity, the solar diameter, the solar neutrino flux, and nuclear decay rates, with implications concerning the Sun’s internal structure and rotation, and neutrino processes,'' \textit{Astroparticle Physics,}} 2013, {\bf 42}, 62-69. \textbf{Topics: P,S,V}

\bibitem{Sturrock BNL} Sturrock P. A., Buncher J. B., Fischbach E., Gruenwald J. T., Javorsek II D., Jenkins J. H., Lee R. H., Matter J. J., Newport J. R., \href{https://doi.org/10.1016/j.astropartphys.2010.06.004}{``Power spectrum analysis of BNL decay rate data,'' \textit{Astroparticle Physics,}} 2010, {\bf 34}, Iss. 2, 121-127. \textbf{Topics: P,S,V}

\bibitem{Sturrock_PTB} Sturrock P. A., Buncher J. B., Fischbach E., Gruenwald J. T., Javorsek II D., Jenkins J. H., Lee R. H., Matter J. J., Newport J. R., \href{https://doi.org/10.1007/s11207-010-9659-4}{``Power Spectrum Analysis of Physikalisch-Technische Bundesanstalt Decay-Rate Data: Evidence for Solar Rotational Modulation,'' \textit{Solar Physics,}} 2010, {\bf 267}, 251-265. \textbf{Topics: P,S,V}

\item Sturrock P.  A., Buncher J. B., Fischbach E., Javorsek II D., Jenkins J. H., Mattes J. J., \href{https://iopscience.iop.org/article/10.1088/0004-637X/737/2/65/meta}{``Concerning the phases of the annual variations of nuclear decay rates,'' \textit{Astrophysical Journal,}} 2011, {\bf 737}, No. 2. \textbf{Topics: A,B,P,V}

\item Sturrock P. A., Caldwell D. O., Scargle J. D., Wheatland M. S., \href{https://journals.aps.org/prd/abstract/10.1103/PhysRevD.72.113004}{``Power-spectrum analyses of Super-Kamiokande solar neutrino data: Variability and its implications for solar physics and neutrino physics,'' \textit{Physical Review D,}} 2005, {\bf 72}, 113004. \textbf{Topics: E,P,S,V}

\item Sturrock P. A., Fischbach E., ``Comparative Analysis of Brookhaven National Laboratory Nuclear Decay Data and Super-Kamiokande Neutrino Data: Indication of a Solar Connection,'' \href{https://arxiv.org/abs/1511.08770}{\underline{arXiv:1511.08770}} [hep-ph], 2015. \textbf{Topics: P,S,V}

\bibitem{Sturrock} Sturrock P. A., Fischbach E., Javorsek II D., Jenkins J.,  Lee R.,``The Case for a Solar Influence on Certain Nuclear Decay Rates,''  \href{https://arxiv.org/abs/1301.3754}{\underline{arXiv:1301.3754}} [hep-ph], 2013. \textbf{Topics: P,S,V}

\bibitem{Sturrock_AstrophysJour} Sturrock P. A., Fischbach E., Jenkins J., \href{https://iopscience.iop.org/article/10.1088/0004-637X/794/1/42}{``Analysis of beta-decay rates for Ag108, Ba133, Eu152, Eu154, Kr85, Ra226, and Sr90, measured at the Physikalisch-Technische Bundesanstalt from 1990-1996,'' \textit{Astrophysical Journal,}} 2014, {\bf 794}, No. 1, 149-155. \textbf{Topics: B,P,V}

\bibitem{Sturrock_Solar_Physics} Sturrock P. A., Fischbach E., Jenkins J., \href{https://doi.org/10.1007/s11207-011-9807-5}{``Further Evidence Suggestive of a Solar Influence on Nuclear Decay Rates,'' \textit{Solar Physics,}} 2011, {\bf 272}, No. 1, 1-10. \textbf{Topics: P,S,V}

\bibitem{Sturrock_2014} Sturrock P. A., Fischbach E., Javorsek II D., Jenkins J. H., Lee R. H., Nistor J., Scargle J. D., \href{https://doi.org/10.1016/j.astropartphys.2014.04.006}{``Comparative study of beta-decay data for eight nuclides measured at the Physikalisch-Technische Bundesanstalt,'' \textit{Astroparticle Physics,}} 2014, {\bf 59}, 47-58. \textbf{Topics: B,P,V}

\item Sturrock P. A., Fischbach E., Parkhomov A., Scargle J. D., Steinitz G., ``Concerning the variability of beta-decay measurements,''  \href{https://arxiv.org/abs/1510.05996}{\underline{arXiv:1510.05996}} [nucl-ex], 2015. \textbf{Topics: B,P,S,V}

\item Sturrock P. A., Fischbach E., Piatibratova O., Steinitz G., Scholkmann F., ``An Oscillation Evident in Both Solar Neutrino Data and Radon Decay Data,'' \href{https://arxiv.org/pdf/1907.11749.pdf}{\underline{arXiv:1907.11749}} [hep-ph], 2019. \textbf{Topics: B,P,S,V}

\item Sturrock P. A., Fischbach E., Scargle J. D., \href{https://link.springer.com/content/pdf/10.1007/s11207-016-1008-9.pdf}{``Comparative Analyses of Brookhaven National Laboratory Nuclear Decay Measurements and Super-Kamiokande Solar Neutrino Measurements: Neutrinos and Neutrino-Induced Beta-Decays as Probes of the Deep Solar Interior,'' \textit{Solar Physics,}} 2016, {\bf 291}, 3467-3484. \textbf{Topics: B,P,S,V}

\item Sturrock P. A., Parkhomov A. G., Fischbach E., Jenkins J. H., \href{https://doi.org/10.1016/j.astropartphys.2012.03.002}{``Power spectrum analysis of LMSU (Lomonosov Moscow State University) nuclear decay-rate data: Further indication of r-mode oscillations in an inner solar tachocline,'' \textit{Astroparticle Physics,}} 2012, {\bf 35}, Iss. 11, 755-758. \textbf{Topics: B,P,S,V}

\item Sturrock P. A., Scargle J. D., Walther G., Wheatland M. S., \href{https://iopscience.iop.org/article/10.1086/312269/meta}{``Rotational Signature and Possible r-Mode Signature in the GALLEX Solar Neutrino Data,'' \textit{Astrophysical Journal Letters,}} 1999, {\bf 523}, No. 2, 177-180. \textbf{Topics: P,S,V}

\item Sturrock P. A., Steinitz G., Fischbach E., \href{https://doi.org/10.1016/j.astropartphys.2018.02.003}{``Analysis of gamma radiation from a radon source. II: Indications of influences of both solar and cosmic neutrinos on beta decays,'' \textit{Astroparticle Physics,}} 2018, {\bf 100}, 1-12. \textbf{Topics: B,P,S,V}

\item Sturrock P. A., Steinitz G., Fischbach E., ``Analysis of Radon-Chain Decay Measurements: Evidence of Solar Influences and Inferences Concerning Solar Internal Structure and the Role of Neutrinos,'' \href{https://arxiv.org/abs/1705.03010}{\underline{arXiv:1705.03010}} [astro-ph.SR], 2017. \textbf{Topics: A,B,T,P,S,V}

\item Sturrock P. A., Steinitz G., Fischbach E., \href{https://doi.org/10.1016/j.astropartphys.2018.01.004}{``Concerning the variability of nuclear decay rates: Rebuttal of an article by Pommé et al.,'' \textit{Astroparticle Physics,}} 2018, {\bf 98}, 9-12. \textbf{Topics: B,E,V}

\bibitem{Sturrock_astroparticle} Sturrock P. A., Steinitz G., Fischbach E., Javorsek II D., Jenkins J. H., \href{https://doi.org/10.1016/j.astropartphys.2012.04.009}{``Analysis of gamma radiation from a radon source: Indications of a solar influence,'' \textit{Astroparticle Physics,}} 2012, {\bf 36}, Iss. 1, 18-25. \textbf{Topics: G,P,S,V}

\item Sturrock P. A., Steinitz G., Fischbach E., Parkhomov A., Scargle J. D., \href{https://doi.org/10.1016/j.astropartphys.2016.07.005}{``Analysis of beta-decay data acquired at the Physikalisch-Technische Bundesanstalt: Evidence of a solar influence,'' \textit{Astroparticle Physics,}} 2016, {\bf 84}, 8-14. \textbf{Topics: B,P,S,V}

\item Sturrock P. A., Walther G., Wheatland M. S., \href{https://iopscience.iop.org/article/10.1086/304955/meta}{``Search for Periodicities in the Homestake Solar Neutrino Data,'' \textit{Astrophysical Journal,}} 1997, {\bf 491}, No. 1, 409-413. \textbf{Topics: P,S,T,V}

\item van Rooy M. W., \href{http://scholar.sun.ac.za/handle/10019.1/97145}{``An investigation of a possible effect of reactor antineutrinos on the decay rate of \textsuperscript{22}Na,''} Ph.D. Thesis, Stellenbosch University, 2015, unpublished. \textbf{Topics: B,S,V}

\item Vasiliev B. V., \href{https://www.scirp.org/pdf/jmp_2020050716255045.pdf}{``The Beta-Decay Induced by Neutrino Flux,'' \textit{Journal of Modern Physics,}} 2020, {\bf 11}, 608-615. \textbf{Topics: B,E,T,V,S}

\item Vasiliev B. V., \href{https://doi.org/10.4236/jmp.2020.111005}{``Effect of Reactor Neutrinos on Beta-Decay,'' \textit{Journal of Modern Physics,}} 2020, {\bf 11}, No. 1, 91-96. \textbf{Topics: B,S,T,V}

\bibitem{Veprev} Veprev D. P., Muromtsev V. I., \href{https://doi.org/10.1016/j.astropartphys.2012.04.012}{``Evidence of solar influence on the tritium decay rate,'' \textit{Astroparticle Physics,}} 2012, {\bf 36}, Iss. 1, 26-30. \textbf{Topics: B,E,S,V}

\item Walg J., Zigel Y., Rodnianski A., Orion I., ``Solar Flare Detection Method using \textsuperscript{222}Rn Radioactive Source,'' \href{https://arxiv.org/abs/2002.02787}{\underline{arXiv:2002.02787}} [astro-ph.SR], 2020. \textbf{Topics: A,E,S,V}

\item Walg J., Zigel Y., Rodnianski A., Orion I., \href{https://www.atiner.gr/presentations/PHY2019-0137.pdf}{``Evidence of Neutrino Flux effect on Alpha Emission Radioactive Half-Life,'' \textit{ATINER Conference Presentation Series,}} 2019, No. PHYS2019-0137. \textbf{Topics: A,E,S,V}

\item Zaqarashvili T. V., Carbonell M., Oliver R., Ballester J. L., \href{https://iopscience.iop.org/article/10.1088/0004-637X/709/2/749/meta}{``Magnetic Rossby waves in the solar Tachocline and Rieger-type Periodicities,'' \textit{Astrophysical Journal,}} 2010, {\bf 709}, No. 2, 749-758. \textbf{Topics: P,T,V}
\newline
\newline
\end{enumerate}

\begin{flushleft}
\large\bf Related Topics
\end{flushleft}

\begin{justify}
The following is a list of references which do not directly discuss the implication of time-dependent variations in radioactive decay rate data or arguments against such variations, but rather provide supporting analysis for the references above.
\end{justify}

\begin{enumerate}
\renewcommand{\labelenumi}{\arabic{enumi}.}

\item Abdurashitov J. N., Gavrin V. N., Girin S. V., Gorbachev V. V., Ibragimova T. V., Kalikhov A. V., Khairnasov N. G., Knodel T. V., Mirmov I. N., Shikhin A. A., SAGE Collaboration, et al., \href{https://doi.org/10.1103/PhysRevC.60.055801}{``Measurement of the solar neutrino capture rate with gallium metal,''  \textit{Physical Review C,}} 1999, {\bf 60}, Iss. 5, 055801. \href{https://arxiv.org/abs/astro-ph/9907113v2}{\underline{arXiv:1902.10131}} [nucl-ex] \textbf{Topics: E,P,S}

\item Aharmim B., Ahmed N., Anthony A. E., Barros N., Beier E. W., Bellerive A., Beltran B., Bergevin M., Biller S. D., Boudjemline K., et al., \href{https://iopscience.iop.org/article/10.1088/0004-637X/710/1/540/meta}{``Searches for high-frequency variations in the \textsuperscript{8}B Solar Neutrino Flux at the Sudbury Neutrino Observatory,''  \textit{Astrophysical Journal,}} 2010, {\bf 710}, No. 1, 540-548. \textbf{Topics: E,L,S}

\item Ahmad I., Bonino G., Castagnoli C., Fischer S. M., Kutschera W., Paul M., \href{https://doi.org/10.1103/PhysRevLett.80.2550}{``Three-Laboratory Measurement of the \textsuperscript{44}Ti Half-life,''  \textit{Physical Review Letters,}} 1998, {\bf 80}, Iss. 12. \textbf{Topics: D,E,L}

\item Al-Bataina B., Janecke J., \href{https://doi.org/10.1524/ract.1987.42.4.159}{``Half-Lives of Long-Lived Alpha Emitters,''  \textit{Radiochimica Acta,}} 1987, {\bf 42}, Iss. 4, 159-164. \textbf{Topics: A,E,P}

\item Aldrich L. T., Wetherill G. W., Tilton G. R., Davis G. L., \href{https://doi.org/10.1103/PhysRev.103.1045}{``Half-Life of Rb\textsuperscript{87},''  \textit{Physical Review,}} 1956, {\bf 103}, Iss. 4, 1045-1047. \textbf{Topics: B,E}

\item Altmann M., Balata M., Belli P., Bellotti E., Bernabei R., Burkert E., Cattadori C., Cerichelli G., Chiarini M., Cribier M., GNO Collaboration, et al., \href{https://doi.org/10.1016/j.physletb.2005.04.068}{``Complete results for five years of GNO solar neutrino observations,''  \textit{Physics Letters B,}} 2005, {\bf 616}, Iss. 3-4, 174-190. \textbf{Topics: L,E,P,S}

\item Altmann M., Balata M., Belli P., Bellotti E., Bernabei R., Burkert E., Cattadori C., Cerichelli G., Chiarini M., Cribier M., GNO Collaboration, et al., \href{https://doi.org/10.1016/S0370-2693(00)00915-1}{``GNO solar neutrino observations: results for GNO I,''  \textit{Physics Letters B,}} 2000, {\bf 490}, Iss. 1-2, 16-26. \textbf{Topics: L,E,P,S}

\item Anderson J. L., Spangler G. W., \href{https://pubs.acs.org/doi/abs/10.1021/j100644a019}{``Serial statistics. Is radioactive decay random,''  \textit{Journal of Physical Chemistry,}} 1973, {\bf 77}, No. 26, 3114-3121. \textbf{Topics: B,P,T}

\item Andreotti E., Arnaboldi C., Avignone III F. T., Balata M., Bandac I., Barucci M., Beeman J. W., Bellini F., Brofferio C.,Bryant A., et al., \href{https://doi.org/10.1016/j.astropartphys.2011.02.002}{``\textsuperscript{130}Te neutrinoless double-beta decay with CUORICINO,''  \textit{Astroparticle Physics,}} 2011, {\bf 34} 822-831. \textbf{Topics: B,L,E}

\item Anselmann P., Fockenbrock R., Hampel W., Heusser G., Kiko J., Kirsten T., Laubenstein M., Pernicka E., Pezzoni S., Rönn U., GALLEX Collaboration, et al., \href{https://doi.org/10.1016/0370-2693(94)01586-2}{``First results from the \textsuperscript{51}Cr neutrino source experiment with the GALLEX detector,''  \textit{Physics Letters B,}} 1995, {\bf 342}, Iss. 1-4, 440-450. \textbf{Topics: L,E,P,S}

\item Anselmann P., Hampel W., Heusser G., Kiko J., Kirsten T., Laubenstein M., Pernicka E., Pezzoni S., Rönn U., Sann M., GALLEX Collaboration, et al., \href{https://s3.cern.ch/inspire-prod-files-5/562c13cd7805a7b58a32ff8a802613cd}{``GALLEX results from the first 30 solar neutrino runs,''  \textit{Physics Letters B,}} 1994, {\bf 327}, Iss. 3-4, 377-385. \textbf{Topics: L,E,S}

\item Anselmann P., Hampel W., Heusser G., Kiko J., Kirsten T., Laubenstein M., Pernicka E., Pezzoni S.,  Plaga R., Rönn U., GALLEX Collaboration, et al., \href{https://doi.org/10.1016/0370-2693(93)91264-N}{``GALLEX solar neutrino observations. The results from GALLEX I and early results from GALLEX II,''  \textit{Physics Letters B,}} 1993, {\bf 314}, Iss. 3-4, 445-458. \textbf{Topics: L,E,S}

\item Anselmann P., Hampel W., Heusser G., Kiko J., Kirsten T., Laubenstein M., Pernicka E., Pezzoni S., Rönn U., Sann M., GALLEX Collaboration, et al., \href{https://doi.org/10.1016/0370-2693(95)00897-T}{``GALLEX solar neutrino observations: complete results for GALLEX II,''  \textit{Physics Letters B,}} 1995, {\bf 357}, Iss. 1-2, 237-247. \textbf{Topics: L,E,S}

\item Anselmann P., Hampel W., Heusser G., Kiko J., Kirsten T., Pernicka E., Plaga R., Rönn U., Sann M., Schlosser C., GALLEX Collaboration, et al., \href{https://doi.org/10.1016/0370-2693(92)91522-B}{``Implications of the GALLEX determination of the solar neutrino flux,''  \textit{Physics Letters B,}} 1992, {\bf 285}, Iss. 4, 390-397. \textbf{Topics: L,E,S}

\item Ashenfelter J., Balantekin B., Baldenegro C. X., Band H. R., Barclay G., Bass C. D., Berish D., Bowden N. S., Bryan C. D., Cherwink J. J., GALLEX Collaboration, et al., \href{https://doi.org/10.1016/j.nima.2015.10.023}{``Background radiation measurements at high power research reactors,''  \textit{Nuclear Instruments and Methods in Physics Research A: Accelerators, Spectrometers, Detectors and Associated Equipment,}} 2016, {\bf 806}, 401-419. \textbf{Topics: L,E}

\item Audi G., Bersillon O., Blachot J., Wapstra A. H., \href{http://hal.in2p3.fr/in2p3-00020241/document}{``The NUBASE evaluation of nuclear and decay properties,''  \textit{Nuclear Physics A,}} 1997, {\bf 624}, 1-122. \textbf{Topics: E,P}

\item Audouze J., Fowler W. A., Schramm D. N., \href{https://doi.org/10.1038/physci238008a0}{``\textsuperscript{176}Lu and s-Process Nucleosynthesis,''  \textit{Nature Physical Science,}} 1972, {\bf 238}, 8-11. \textbf{Topics: E,P}

\item Bahcall J. N., \href{https://doi.org/10.1103/PhysRevLett.61.2650}{`Solar flares and neutrino detectors,''  \textit{Physical Review,}} 1988, {\bf 61}, Iss. 23, 2650-2652. \textbf{Topics: P,S}

\item Bahcall J. N., Pinsonneault M. H., Basu S., \href{https://iopscience.iop.org/article/10.1086/321493/meta}{''Solar Models: Current Epoch and Time Dependences, Neutrinos, and Helioseismological Properties,''  \textit{Astrophysical Journal,}} 2001, {\bf 555}, No. 2, 990-1012. \textbf{Topics: P,S}

\item Barabash A. S., \href{https://doi.org/10.1016/j.nuclphysa.2015.01.001}{``Average and recommended half-life values for two-neutrino double beta decay,''  \textit{Nuclear Physics A,}} 2015, {\bf 935}, 52-64. \textbf{Topics: B,P,R,S}

\item Baranyai A., Hirn A., Deme S., Csöke A., Apáthy I., Fehér I., \href{https://doi.org/10.1016/j.apradiso.2019.109028}{``\textsuperscript{212}Bi-\textsuperscript{212}Po alpha source for the calibration and functional testing of silicon detectors: Preparation and characterisation,''  \textit{Applied Radiation and Isotopes,}} 2020, {\bf 158}, 109028. \textbf{Topics: A,E}

\item Beard G. B., Kelly W. H., \href{https://doi.org/10.1016/0029-5582(58)90149-4}{``The use of a samarium loaded liquid scintillator for the determination of the half-life of Sm\textsuperscript{147},''  \textit{Nuclear Physics,}} 1958, {\bf 8}, 207-209. \textbf{Topics: A,E}

\item Beard G. B., Wiedenbeck M. L., \href{https://doi.org/10.1103/PhysRev.95.1245}{``Natural Radioactivity of Sm\textsuperscript{147},''  \textit{Physical Review,}} 1954, {\bf 95}, Iss. 5, 1245-1246. \textbf{Topics: A,E}

\item Beckinsale R. D., Gale N. H., \href{https://doi.org/10.1016/0012-821X(69)90170-8}{``A reappraisal of the decay constants and branching ratio of \textsuperscript{40}K,''  \textit{Earth and Planetary Science Letters,}} 1969, {\bf 6}, Iss. 4, 289-294. \textbf{Topics: B,P}

\item Becquerel H., \href{https://inspirehep.net/literature/47893}{``On the rays emitted by phosphorescence,''  \textit{Compt. Rend. Hebd. Seances Acad. Sci.,}} 1896, {\bf 122}, 420-421. \textbf{Topics: E}

\item Begemann F., Ludwig K. R., Lugmair G. W., Min K., Nyquist L. E., Patchett P. J., Renne P. R., Shih C. Y., Villa I. M., Walker R. J., \href{https://doi.org/10.1016/S0016-7037(00)00512-3}{``Call for an improved set of decay constants for geochronological use,''  \textit{Geochimica et Cosmochimica Acta,}} 2001, {\bf 65}, Iss. 1, 111-121. \textbf{Topics: D,P}

\item Bell R. E., Sosniak J., \href{https://doi.org/10.1139/p59-001}{``Genetic Measurement of the Half life of Bi\textsuperscript{207},''  \textit{Canadian Journal of Physics,}} 1959, {\bf 37}, No. 1. \textbf{Topics: D,EC}

\item Belli P., Bernabei R., Cappella F., Cerulli R., Danevich F. A., Dubovik A. M., d'Angelo S., Galashov E. N., Grinyov B. V., Incicchitti A., et al., \href{https://doi.org/10.1016/j.nima.2010.10.027}{``Radioactive contamination of ZnWO\textsubscript{4} crystal scintillators,''  \textit{Nuclear Instruments and Methods in Physics Research Section A: Accelerators, Spectrometers, Detectors and Associated Equipment,}} 2011, {\bf 626-627}, 31-38. \textbf{Topics: A,B,E,P}

\item Bellini G., Benziger J., Bick D., Bonetti S., Bonfini G., Buizza Avanzini M., Caccianiga B., Cadonati L., Calaprice F., Carraro C., Borexino Collaboration, et al., \href{https://doi.org/10.1016/j.physletb.2011.11.025}{``Absence of a day–night asymmetry in the \textsuperscript{7}Be solar neutrino rate in Borexino,''  \textit{Physics Letters B,}} 2012, {\bf 707}, Iss. 1, 22-26. \textbf{Topics: A,E,P,S}

\item Bellini G., Benziger J., Bonetti S., Buizza Avanzini M., Caccianiga B., Cadonati L., Calaprice F., Carraro C., Chavarria A., Chepurnov A., Borexino Collaboration, et al., \href{https://doi.org/10.1103/PhysRevD.82.033006}{``Measurement of the solar \textsuperscript{8}B neutrino rate with a liquid scintillator target and 3MeV energy threshold in the Borexino detector,''  \textit{Physics Review D,}} 2010, {\bf 82}, 033006. \href{https://arxiv.org/abs/0808.2868v3}{\underline{arXiv:0808.2868}} [astro-ph] \textbf{Topics: E,P,S}

\item Bieber J. W., Seckel D., Stanev T., Steigman G., \href{https://www.nature.com/articles/348407a0.pdf?origin=ppub}{``Variation of the solar neutrino flux with the Sun's activity,''  \textit{Nature,}} 1990, {\bf 348}, No. 6300, 407-411. \textbf{Topics: P,S}

\item Bizzeti P. G., Bizzeti-Sona A. M., \href{https://doi.org/10.1016/j.nima.2013.11.017}{``Integral- and differential-decay-curve methods for recoil distance measurements of nuclear lifetimes,''  \textit{Nuclear Instruments and Methods in Physics Research Section A: Accelerators, Spectrometers, Detectors and Associated Equipment,}} 2014, {\bf 736}, 179-183. \textbf{Topics: P}

\item Brancaccio F., Dias M. S., Koskinas M. F., Moreira D. S., de Toledo F., \href{https://doi.org/10.1016/j.apradiso.2019.108900}{``Data analysis software package for radionuclide standardization with a digital coincidence counting system, \textit{Applied Radiation and Isotopes,}} 2020, {\bf 155}, 108900. \textbf{Topics: E,P}

\item Brinkman G. A., Aten Jr. A. H. W., Veenboer J. T., \href{https://doi.org/10.1016/0031-8914(65)90056-X}{``Natural radioactivity of K-40, Rb-87, and Lu-176,'' \textit{Physica,}} 1965, {\bf 31}, Iss 8, 1305-1319. \textbf{Topics: B,E,P}

\item Budick B., Chen J., Lin H., \href{https://doi.org/10.1103/PhysRevLett.67.2630}{``Half-Life of Molecular Tritium and the Axial-Vector Interaction in Tritium \textbeta Decay,'' \textit{Physical Review Letters,}} 1991, {\bf 67}, Iss. 19, 2630. \textbf{Topics: B,D,P}

\item Byun J., Hwang H., Yun J., \href{https://doi.org/10.1016/j.apradiso.2019.108932}{``A low background gamma-ray spectrometer with a large well HPGe detector,'' \textit{Applied Radiation and Isotopes,}} 2020, {\bf 156}, 1080932. \textbf{Topics: B,T}

\item Caldwell D. O., Sturrock P. A., \href{https://doi.org/10.1016/j.astropartphys.2005.05.001}{``Evidence for solar neutrino flux variability and its implications,'' \textit{Astroparticle Physics,}} 2005, {\bf 23}, Iss. 6, 543-556. \textbf{Topics: P,S}

\item Casanovas R., Morant J. J., Salvadó M., \href{https://doi.org/10.1016/j.radmeas.2012.06.001}{``Temperature peak-shift correction methods for NaI(Tl) and LaBr\textsubscript{3}(Ce) gamma-ray spectrum stabilisation,'' \textit{Radiation Measurements,}} 2012, {\bf 47}, Iss. 8, 588-595. \textbf{Topics: E,P}

\item Chen Y., Kashy E., Bazin D., Benenson W., Morrissey D. J., Orr N. A., Sherrill B. M., Winger J. A., Young B., Yurkon J., \href{https://doi.org/10.1103/PhysRevC.47.1462}{``Half-life of \textsuperscript{32}Si,'' \textit{Physical Review C,}} 1993, {\bf 47,} Iss. 4. \textbf{Topics: B,E}

\item Cheng C., Wei Z., Hei D., Jia W., Li J., Cai P., Gao Y., Shan Q., Ling Y., \href{https://doi.org/10.1016/j.apradiso.2020.109045}{``MCNP benchmark of a \textsuperscript{252}Cf source-based PGNAA system for bulk sample analysis,'' \textit{Applied Radiation and Isotopes,}} 2020, {\bf 158}, 109045. \textbf{Topics: A,E,T}

\item Collins S. M., Shearman R., Ivanov P., Regan P. H., \href{https://doi.org/10.1016/j.apradiso.2019.109021}{``The impact of high-energy tailing in high-purity germanium gamma-ray spectrometry on the activity determination of \textsuperscript{224}Ra using the 241.0 keV emission,'' \textit{Applied Radiation and Isotopes,}} 2020, {\bf 157}, 109021. \textbf{Topics: A,E,P}

\item da Silva M. A. L., Poledna R., Iwahara A., da Silva C. J., Delgado J. U., Lopes R. T., \href{https://doi.org/10.1016/j.apradiso.2006.02.058}{``Standardization and decay data determinations of \textsuperscript{125}I, \textsuperscript{54}Mn and \textsuperscript{203}Hg,'' \textit{Applied Radiation and Isotopes,}} 2006, {\bf 64}, Iss. 10, 1440-1445. \textbf{Topics: D,E}

\item Das C. R., Pulido J., Picariello M., \href{https://journals.aps.org/prd/abstract/10.1103/PhysRevD.79.073010}{``Light sterile neutrinos, spin flavor precession, and the solar neutrino experiments,'' \textit{Physical Review D,}} 2009, {\bf 79}, Iss. 7, 073010. \textbf{Topics: E,P,S}

\item Davis Jr. R., \href{https://doi.org/10.1016/0920-5632(96)00263-0}{``A review of measurements of the solar neutrino flux and their variation,'' \textit{Nuclear Physics B- Proceedings Supplements,}} 1996, {\bf 48}, Iss. 1-3, 284-298. \textbf{Topics: R,S}

\item De Bièvre P., Verbruggen A., \href{https://doi.org/10.1088/0026-1394/36/1/5}{``A new measurement of the half-life of \textsuperscript{241}Pu using isotope mass spectrometry,'' \textit{Metrologia}} 1999, {\bf 36}, No.1, 25-31. \textbf{Topics: B,D,E,P}

\item De Ruyter A. W., Aten Jr A. H. W., Van Dulmen A., Krol-Koning C., Zuidema E., \href{https://doi.org/10.1016/0031-8914(66)90029-2}{``Specific gamma emission of natural potassium and lanthanum,'' \textit{Physica,}} 1966, {\bf 32}, Iss. 5, 991-994. \textbf{Topics: B,E,P}

\item Delache P., Gavryusev V., Gavryuseva E., Laclare F., Régulo C., Roca Cortés T., \href{http://adsabs.harvard.edu/full/1993ApJ...407..801D}{``Time correlation between solar structural parameters - p-mode frequencies, radius, and neutrino flux,'' \textit{Astrophysical Journal,}} 1993, {\bf 407}, No. 2, 801-805. \textbf{Topics: E,T,S}

\item Dryák P., Šolc J., \href{https://doi.org/10.1016/j.apradiso.2019.108942}{``Measurement of the branching ratio related to the internal pair production of Y-90,'' \textit{Applied Radiation and Isotopes,}} 2020, {\bf 156}, 108942. \textbf{Topics: B,E,P}

\item Elmore D., Anantaraman N., Fulbright H. W., Gove H. E., Hans H. S.,Nishiizumi K., Murrell M. T., Honda M., \href{https://doi.org/10.1103/PhysRevLett.45.589}{``Half-Life of \textsuperscript{32}Si from Tandem-Accelerator Mass Spectrometry,'' \textit{Physical Review Letters,}} 1980, {\bf 45}, Iss. 8. \textbf{Topics: B,D}

\item Frekers D., Ejiri H., Akimune H., Adachi T., Bilgier B., Brown B. A., Cleveland B. T., Fujita H., Fujiware M., Ganioglu E., et al., \href{https://doi.org/10.1016/j.physletb.2011.10.061}{``The \textsuperscript{71}Ga(\textsuperscript{3}He,t) reaction and the low-energy neutrino response,'' \textit{Physics Letters B,}} 2011, {\bf 706}, Iss. 2-3, 134-138. \textbf{Topics: B,E,P,S}

\item García-Toraño E., Crespo T., Marouli M., Jobbágy V., Pommé S., Ivanov P., \href{https://doi.org/10.1016/j.apradiso.2019.108863}{``Alpha-particle emission probabilities of \textsuperscript{231}Pa derived from first semiconductor spectrometric measurements,'' \textit{Applied Radiation and Isotopes,}} 2019, {\bf 154}, 108863. \textbf{Topics: A,E}

\item Görres J., Meibner J., Schatz H., Stech E., Tischhauser P., Wiescher M., Bazin D., Harkewicz R., Hellström M., Sherrill B., et al., \href{https://doi.org/10.1103/PhysRevLett.80.2554}{``Half-Life of \textsuperscript{44}Ti as a Probe for Supernova Models,'' \textit{Physical Review Letters,}} 1998, {\bf 80}, Iss. 12, 2554. \textbf{Topics: D}

\item Hampel W., Handt J., Heusser G., Kiko J., Kirsten T., Laubenstein M., Pernicka E., Rau W., Wojcik M., Zakharov Y., GALLEX Collaboration, et al., \href{https://doi.org/10.1016/S0370-2693(98)01579-2}{``GALLEX solar neutrino observations: results for GALLEX IV,'' \textit{Physics Letters B,}} 1999, {\bf 447}, Iss. 1-2, 127-133. \textbf{Topics: E,S}

\item Hampel W., Heusser G., Kiko J., Kirsten T., Laubenstein M., Pernicka E., Rau W., Rönn U., Schlosser C., Wojcik M., GALLEX Collaboration, et al., \href{https://doi.org/10.1016/S0370-2693(97)01562-1}{``Final results of the \textsuperscript{51}Cr neutrino source experiments in GALLEX,'' \textit{Physics Letters B,}} 1998, {\bf 420}, Iss. 1-2, 114-126. \textbf{Topics: E,S}

\item Hampel W., Heusser G., Kiko J., Kirsten T., Laubenstein M., Pernicka E., Rau W., Rönn U., Schlosser C., Wojcik M., GALLEX Collaboration, et al., \href{https://doi.org/10.1016/S0370-2693(96)01121-5}{``GALLEX solar neutrino observations: Results for GALLEX III,'' \textit{Physics Letters B,}} 1996, {\bf 388}, Iss. 2, 384-396. \textbf{Topics: E,S}

\item Haubold H. J., Gerth E., \href{https://doi.org/10.1007/BF00152173}{``On the Fourier spectrum analysis of the solar neutrino capture rate,'' \textit{Solar Physics,}} 1990, {\bf 127}, 347-356. \textbf{Topics: E,P,S,T}

\item Haubold H. J., Gerth E., \href{https://doi.org/10.1007/BF00653522}{``Time variations of the solar neutrino flux,'' \textit{Astrophysics and Space Science,}} 1985, {\bf 112}, 397-405. \textbf{Topics: P,S,T}

\item Horvat V., Hardy J. C., \href{https://doi.org/10.1016/j.nima.2013.02.047}{``Time-interval analysis of beta decay,'' \textit{Nuclear Instruments and Methods in Physics Research Section A: Accelerators, Spectrometers, Detectors and Associated Equipment,}} 2013, {\bf 713}, 19-26. \textbf{Topics: B,P}

\item Jaffey A. H., \href{https://doi.org/10.1016/0029-554X(70)90624-5}{``Statistical considerations in half-life measurement, I,'' \textit{Nuclear Instruments and Methods,}} 1970, {\bf 81}, Iss. 1, 155-163. \textbf{Topics: P,T}

\item Kaether F., Hampel W., Heusser G., Kiko J., Kirsten T., \href{https://doi.org/10.1016/j.physletb.2010.01.030}{``Reanalysis of the Gallex solar neutrino flux and source experiments,'' \textit{Physics Letters B,}} 2010, {\bf 685}, Iss. 1, 47-54. \textbf{Topics: P,S}

\item Kelly W. H., Beard G. B., Peters R. A., \href{https://doi.org/10.1016/0029-5582(59)90292-5}{``The beta decay of K\textsuperscript{40},'' \textit{Nuclear Physics,}} 1959, {\bf 11}, 492-498. \textbf{Topics: B,E}

\item Kirsten T. A., \href{https://doi.org/10.1007/BF02508123}{``Results from solar-neutrino experiments,'' \textit{II Nuovo Cimento C,}} 1996, {\bf 19}, No. 6, 821-833. \textbf{Topics: E,P,S}

\item Kirsten T. A., GNO Collaboration, \href{https://doi.org/10.1016/S0920-5632(03)01301-X}{``Progress in GNO,'' \textit{Nuclear Physics B - Proceedings Supplements,}} 2003, {\bf 118}, 33-38. \textbf{Topics: E,EC,L,P,S}

\item Kossert K., Nähle O. J.,  Carles A. G., \href{https://doi.org/10.1016/j.apradiso.2011.03.046}{``Beta shape-factor function and activity determination \textsuperscript{241}Pu,'' \textit{Applied Radiation and Isotopes,}} 2011, {\bf 69}, Iss. 9, 1246-1250. \textbf{Topics: B,E,P}

\item Lean J. L., Bruecknew G. E., \href{http://adsabs.harvard.edu/full/1989ApJ...337..568L}{``Intermediate-term solar periodicities - 100-500 days,'' \textit{Astrophysical Journal,}} 1989, {\bf 337}, 568-578. \textbf{Topics: E,P,S}

\item Li Y. F., Xing Z. Z., \href{https://doi.org/10.1016/j.physletb.2010.11.015}{``Possible capture of KeV sterile neutrino dark matter on radioactive β-decaying nuclei,'' \textit{Physics Letters B,}} 2011, {\bf 695}, Iss. 1-4, 205-210. \textbf{Topics: B,P,S,T}

\item Liu H., Zhou Q., Fan F., Liang J., Zhang M., \href{https://doi.org/10.1016/j.apradiso.2019.108944}{``Activity determination of \textsuperscript{231}Pa by means of liquid scintillation counting,'' \textit{Applied Radiation and Isotopes,}} 2020, {\bf 155}, 108944. \textbf{Topics: A,B,E,P}

\item Lomb N. R., \href{https://doi.org/10.1007/BF00648343}{``Least-squares frequency analysis of unequally spaced data,'' \textit{Astrophysics and Space Science,}} 1976, {\bf 39}, 447-462. \textbf{Topics: P,T}

\item Luca A., \href{https://doi.org/10.1016/j.apradiso.2019.108941}{``\textsuperscript{226}Th nuclear decay data evaluation,'' \textit{Applied Radiation and Isotopes,}} 2020, {\bf 155}, 108941. \textbf{Topics: A,E,P}

\item MacMahon D., Pearce A., Harris P., \href{https://doi.org/10.1016/j.apradiso.2003.11.028}{``Convergence of techniques for the evaluation of discrepant data,'' \textit{Applied Radiation and Isotopes,}} 2004, {\bf 60}, Iss. 2-4, 275-281. \textbf{Topics: D.P}

\item Marouli M., Lutter G., Pommé S., Van Ammel R., Hult M., Pierre S., Dryák P., Carconi P., Fazio A., Brunchertseifer F., et al., \href{https://doi.org/10.1016/j.apradiso.2018.08.023}{``Measurement of absolute γ-ray emission probabilities in the decay of \textsuperscript{227}Ac in equilibrium with its progeny,'' \textit{Applied Radiation and Isotopes,}} 2019, {\bf 144}, 34-46. \textbf{Topics: A,E,P}

\item Marouli M., Suliman G., Pommé S., Van Ammel R., Jobbágy V., Stroh H., Dikmen H., Paepen J., Dirican A., Bruchertseifer F., et al., \href{https://doi.org/10.1016/j.apradiso.2012.12.005}{``Decay data measurements on \textsuperscript{213}Bi using recoil atoms,'' \textit{Applied Radiation and Isotopes,}} 2013, {\bf 74}, 123-127. \textbf{Topics: A,B,E,P}

\item Marouli M., Pommé S., \href{https://doi.org/10.1016/j.apradiso.2019.108821}{``Automated optical distance measurements for counting at a defined solid angle,'' \textit{Applied Radiation and Isotopes,}} 2019, {\bf 153}, 108821. \textbf{Topics: A,E,P}

\item Massetti S., Stroini M., \href{https://iopscience.iop.org/article/10.1086/178112}{``Spacetime Modulation of Solar Neutrino Flux: 1970-1992,'' \textit{Astrophysical Journal,}} 1996, {\bf 472}, No. 2, 827-831. \textbf{Topics: P,S}

\item McNutt Jr. R. L., \href{https://search.proquest.com/docview/213560723?accountid=13360}{``Correlated Variations in the Solar Neutrino Flux and the Solar Wind and the Relation to the Solar Neutrino Problem,'' \textit{Science,}} 1995, {\bf 270}, Iss. 5242, 1635-1639. \textbf{Topics: P,S}

\item Aker M., Altenmüller K., Arenz M., Babutzka M., Barrett J., Bauer S., Beck M., Beglarian A., Behrens J., Bergmann T., KATRIN collaboration, \href{https://journals.aps.org/prl/pdf/10.1103/PhysRevLett.123.221802}{``Improved Upper Limit on Neutrino Mass from a Direct Kinematic Method by KATRIN,'' \textit{Physical Review Letters,}} 2019, {\bf 123}, 221802. \textbf{Topics: B,E,P,S}

\item Mitra M., Senjanović G., Vissani F., \href{https://doi.org/10.1016/j.nuclphysb.2011.10.035}{``Neutrinoless double beta decay and heavy sterileneutrinos,'' \textit{Nuclear Physics B,}} 2012, {\bf 856}, Iss. 1, 26-73. \textbf{Topics: B,S,T}

\item Napoli E., Cessna J. T., Fitzgerald R., Pibida L., Collé R., Laureano-Pérez L., Zimmerman B. E., Bergeron D. E., \href{https://doi.org/10.1016/j.apradiso.2019.108933}{``Primary standardization of \textsuperscript{224}Ra activity by liquid scintillation counting,'' \textit{Applied Radiation and Isotopes,}} 2020, {\bf 155}, 108933. \textbf{Topics: A,B,E,P}

\item Oakley D. S., Snodgrass, H. B., Ulrich, R. K., Vandekop, T. L.,  \href{http://adsabs.harvard.edu/full/1994ApJ...437L..63O}{``On the correlation of solar surface magnetic flux with solar neutrino capture rate,'' \textit{Astrophysical Journal,}} 1994, {\bf 437}, L63-L66. \textbf{Topics: E,P,S}

\item Pandola L., \href{https://doi.org/10.1016/j.astropartphys.2004.07.007}{``Search for time modulations in the Gallex/GNO solar neutrino data,'' \textit{Astroparticle Physics,}} 2004, {\bf 22}, Iss. 2, 219-226. \textbf{Topics: E,P,S}

\item Pibida L., Zimmerman B. E., King L., Fitzgerald R., Bergeron D. E., Napoli E., Cessna J. T., \href{https://doi.org/10.1016/j.apradiso.2019.108943}{``Determination of the internal pair production branching ratio of \textsuperscript{90}Y,'' \textit{Applied Radiation and Isotopes,}} 2020, {\bf 156}, 108943. \textbf{Topics: B,E}

\item Picoreti R., Guzzo M. M., de Holanda P. C., Peres O. L. G., \href{https://doi.org/10.1016/j.physletb.2016.08.007}{``Neutrino decay and solar neutrino seasonal effect,'' \textit{Physics Letters B,}} 2016, {\bf 761}, 70-73. \textbf{Topics: P,S,T}

\item Pierre S., Cassette P., Loidl M., Branger T., Lacour D., Le Garrérès I., Morelli S., \href{https://doi.org/10.1016/j.apradiso.2009.11.035}{``On the variation of the \textsuperscript{210}Po half-life at low temperature,'' \textit{Applied Radiation and Isotopes,}} 2010, {\bf 68}, Iss. 7-8, 1467-1470. \textbf{Topics: E}

\item Pommé S., \href{https://doi.org/10.1016/j.apradiso.2008.02.038} {``Cascades of pile-up and dead time,'' \textit{Applied Radiation and Isotopes,}} 2008, {\bf 66}, Iss. 6-7, 941-947. \textbf{Topics: P,T}

\item Pommé S., \href{https://pubs.acs.org/doi/abs/10.1021/bk-2007-0945.ch016} {``Dead Time, Pile-Up, and Counting Statistics,'' \textit{Applied Modeling and Computations in Nuclear Science,}} 2006, {\bf 945}, Ch. 16, 218-233. \textbf{Topics: P,T}

\item Pommé S., \href{https://iopscience.iop.org/article/10.1088/0026-1394/52/3/S73/meta} {``The uncertainty of counting at a defined solid angle,'' \textit{Metrologia,}} 2015, {\bf 52}, No. 3, S73-S85. \textbf{Topics: P,T}

\item Pommé S., \href{https://iopscience-iop-org.ezproxy.lib.purdue.edu/article/10.1088/0026-1394/53/2/S55} {``When the model doesn't cover reality: examples from radionuclide metrology,'' \textit{Metrologia,}} 2016, {\bf 53}, S55-S64. \textbf{Topics: P,T}

\item Pommé S., Denecke B., Alzetta J. P., \href{https://doi.org/10.1016/S0168-9002(99)00016-9} {``Influence of pileup rejection on nuclear counting, viewed from the time-domain perspective,'' \textit{Nuclear Instruments and Methods in Physics Research Section A: Accelerators, Spectrometers, Detectors and Associated Equipment,}} 1999, {\bf 426}, Iss. 2-3, 564-582. \textbf{Topics: E,T}

\item Pommé S., Fitzgerald R., Keightley J., \href{https://iopscience.iop.org/article/10.1088/0026-1394/52/3/S3/meta} {``Uncertainty of nuclear counting,'' \textit{Metrologia,}} 2015, {\bf 52}, No. 3, S3-S17. \textbf{Topics: P,T}

\item Pommé S., Keightley J. \href{https://pubs.acs.org/doi/abs/10.1021/bk-2007-0945.ch023}{``Count Rate Estimation of a Poisson Process: Unbiased Fit versus Central Moment Analysis of Time Interval Spectra,'' \textit{Applied Modeling and Computations in Nuclear Science,}} 2006, {\bf 945}, Ch. 23, 316-334. \textbf{Topics: P}

\item Pommé S., Keightley J. \href{https://iopscience.iop.org/article/10.1088/0026-1394/52/3/S200/meta}{``Determination of a reference value and its uncertainty through a power-moderated mean,'' \textit{Metrologia,}} 2015, {\bf 52}, No. 3, S200-S212. \textbf{Topics: P}

\item Pommé S., Paepen J., Marouli M., \href{https://doi.org/10.1016/j.apradiso.2019.108848} {``Conversion electron spectroscopy of the 59.54 keV transition in \textsuperscript{241}Am alpha decay,'' \textit{Applied Radiation and Isotopes,}} 2019, {\bf 153}, 108848. \textbf{Topics: A,P}

\item Quarati F. G. A., Khodyuk I. V., van Eijk C. W. E., Quarati P., Dorenbos P., \href{https://doi.org/10.1016/j.nima.2012.04.066}{`Study of \textsuperscript{138}La radioactive decays using LaBr\textsubscript{3} scintillators,'' \textit{Nuclear Instruments and Methods in Physics Research Section A: Accelerators, Spectrometers, Detectors and Associated Equipment,}} 2012, {\bf 683}, 46-52. \textbf{Topics: B,E,EC}

\item Ranucci G., \href{https://journals.aps.org/prd/abstract/10.1103/PhysRevD.73.103003}{``Likelihood scan of the Super-Kamiokande I time series data,'' \textit{Physical Review D,}} 2006, {\bf 73}, Iss. 1, 103003. \textbf{Topics: P,T,S}

\item Ranucci G., Rovere M., \href{http://dx.doi.org/10.1103/PhysRevD.75.013010}{`Periodogram and likelihood periodicity search in the SNO solar neutrino data,'' \textit{Physical Review D,}} 2007, {\bf 75}, Iss. 1, 013010. \textbf{Topics: P,T,S}

\item Ratel G., \href{https://iopscience.iop.org/article/10.1088/0026-1394/44/4/S02}{`The Système International de Référence and its application in key comparisons,'' \textit{Metrologia,}} 2007, {\bf 44}, S7-S16. \textbf{Topics: P}

\item Rieger E., Share G. H. Forrest D. J. Kanback G., Reppin C., Chupp E. L., \href{https://doi.org/10.1038/312623a0}{``A 154-day periodicity in the occurrence of hard solar flares?,'' \textit{Nature,}} 1984, {\bf 312}, 623-625. \textbf{Topics: E,S}

\item Rothe C., Hintschich S. I., Monkman A. P., \href{https://www.researchgate.net/profile/Andy_Monkman/publication/7069976_Violation_of_the_Exponential-Decay_Law_at_Long_Times/links/00b4952bf5ed9652e1000000.pdf}{`Violation of the Exponential-Decay Law at Long Times,'' \textit{Physical Review,}} 2006, {\bf 96}, 163601. \textbf{Topics: P,T}

\item Rutherford, E., \href{https://play.google.com/store/books/details?id=UeEdAAAAMAAJ&rdid=book-UeEdAAAAMAAJ&rdot=1}{``Radioactive Substances and their Radiations,'' \textit{Cambridge University Press,}} 1913, 505-507 \textbf{Topics: E}

\item Rutherford, E., Chadwick J., Ellis C. D., \href{https://rb.gy/pk2to5}{``Radiations from Radioactive Substances,'' \textit{Cambridge University Press,}} 1951. \textbf{Topics: E}

\item Rutherford, E., Soddy F., \href{https://doi.org/10.1080/14786440309462960}{``LX. Radioactive Change,'' \textit{The London, Edinburgh, and Dublin Philosophical Magazine and Journal of Science,}} 1903, {\bf 5}, Iss. 29, 576-591. \textbf{Topics: E}

\item Ryazhskaya O. G., Volkova L. V., Zatsepin G. T., \href{https://doi.org/10.1016/S0920-5632(02)01509-8}{``Neutrinos from solar flares at the earth,'' \textit{Nuclear Physics B- Proceedings Supplements,}} 2002, {\bf 110}, 358-360. \textbf{Topics: S}

\item Saio H., \href{http://adsabs.harvard.edu/full/1982ApJ...256..717S}{``R-mode oscillations in uniformly rotating stars,'' \textit{Astrophysical Journal,}} 1982, {\bf 256}, 717-735. \textbf{Topics: T}

\item Sakurai K., \href{https://doi.org/10.1038/278146a0}{``Quasi-biennial variation of the solar neutrino flux and solar activity,'' \textit{Nature,}} 1979, {\bf 278}, 146-148. \textbf{Topics: P,S}

\item Scargle J. D., \href{http://articles.adsabs.harvard.edu/full/1982ApJ...263..835S}{``Studies in astronomical time series analysis. II - Statistical aspects of spectral analysis of unevenly spaced data,'' \textit{Astrophysical Journal,}} 1982, {\bf 263}, 835-853 \textbf{Topics: P}

\item Schön R., Winkler G., Kutschera W.,  \href{https://doi.org/10.1016/j.apradiso.2003.11.027}{``A critical review of experimental data for the half-lives of the uranium isotopes \textsuperscript{238}U and \textsuperscript{235}U,'' \textit{Applied Radiation and Isotopes,}} 2004, {\bf 60}, Iss. 2-4, 263-273. \textbf{Topics: D,P}

\item Schou J., Howe R., Basu S., Christensen-Dalsgaard J., Corbard T., Hill F., Komm R., Larsen R. M., Rabello-Soares M. C., Thompson M. J., \href{https://iopscience.iop.org/article/10.1086/338665/pdf}{``A comparison of solar p-mode parameters from the Michelson doppler imager and the global oscillation network group: Splitting coefficients and rotation inversions,'' \textit{Astrophysical Journal,}} 2002, {\bf 567}, 1234-1249. \textbf{Topics: E,P}

\item Schulc M., Košt'ál M., Šimon J., Novák E., Mareček M., Kubín R., \href{https://doi.org/10.1016/j.apradiso.2019.108937}{``Validation of IRDFF-II library by means of \textsuperscript{252}Cf spectral averaged cross sections,'' \textit{Applied Radiation and Isotopes,}} 2020, {\bf 155}, 108937. \textbf{Topics: E,A}

\item Silverman M. P., \href{https://iopscience.iop.org/article/10.1209/0295-5075/105/22001/meta?casa_token=cOJ-i52ASiwAAAAA:Ph5_O2Ps5xswxG6veSeCgzr_pwgAhsLgHBCqtXyblVQzeZDcS4-sDxd9nhJ81yWoMutILWDpv-XcTOfDHFQ3}{``Theory of nuclear half-life determination by statistical sampling,'' \textit{Europhysics Letters,}} 2013, {\bf 105}, No. 2, 22001. \textbf{Topics: T,P,B,EC}

\item Sima O., De Vismes Ott A., Dias M. S., Dryak P., Ferreux L., Gurau D., Hurtado S., Jodlowski P., Karfopoulos K., Koskinas M. F., et al., \href{https://doi.org/10.1016/j.apradiso.2019.108921}{``Consistency test of coincidence-summing calculation methods for extended sources,'' \textit{Applied Radiation and Isotopes,}} 2020, {\bf 155}, 108921. \textbf{Topics: T,P,B,EC}

\item Smy M. B., Super-Kamiokande Collaboration \href{https://doi.org/10.1016/j.nuclphysbps.2015.06.035}{``Terrestrial Matter Effects from Solar Neutrino Interactions in Super-Kamiokande,'' \textit{Nuclear and Particle Physics Proceedings,}} 2015, {\bf 265-266}, 135-138. \textbf{Topics: E,S}

\item Sturrock P. A., \href{https://iopscience.iop.org/article/10.1086/382141/meta}{Analysis of Super-Kamiokande 5 Day Measurements of the Solar Neutrino Flux,'' \textit{Astrophysical Journal,}} 2004, {\bf 605}, No. 1, 568-572. \textbf{Topics: P,T,S}

\item Sturrock P. A., \href{https://link.springer.com/content/pdf/10.1007/s11207-008-9253-1.pdf}{Evidence for r-Mode Oscillations in Super-Kamiokande Solar Neutrino Data,'' \textit{Solar Physics,}} 2008, {\bf 252}, 221-233. \textbf{Topics: P,T,S}

\item Sturrock P. A., \href{https://iopscience.iop.org/article/10.1086/594993/meta}{"Solar Neutrino Variability and Its Implications for Solar Physics and Neutrino Physics,'' \textit{Astrophysical Journal Letters,}} 2008, {\bf 688}, No. 1, L53-L56. \textbf{Topics: P,T,S}

\item Sturrock P. A., \href{https://doi.org/10.1007/s11207-008-9254-0}{"Time-Frequency Analysis of GALLEX and GNO Solar Neutrino Data,'' \textit{Solar Physics,}} 2008, {\bf 252}, 1-18. \textbf{Topics: P,T,S}

\item Sturrock P. A., Scargle J. D.,  \href{https://doi.org/10.1007/s11207-006-0143-0}{``Power-Spectrum Analysis of Super-Kamiokande Solar Neutrino Data, Taking into Account Asymmetry in the Error Estimates,'' \textit{Solar Physics,}} 2006, {\bf 237}, 1-11. \textbf{Topics: P,S}

\item Sturrock P. A., Walther G., Wheatland M. S.,  \href{https://iopscience.iop.org/article/10.1086/306353/pdf}{``Apparent Latitudinal Modulation of the Solar Neutrino Flux,'' \textit{Astrophysical Journal,}} 1998, {\bf 507}, 978-983. \textbf{Topics: P,T,S}

\item Tilley D. R., Madansky L., \href{https://doi.org/10.1103/PhysRev.116.413}{``Search for Positron Emission in K\textsuperscript{40},'' \textit{Physical Review,}} 1959, {\bf 116}, 413. \textbf{Topics: E,B}

\item Unterweger M. P., \href{https://doi.org/10.1016/S0969-8043(01)00177-4}{``Half-life measurements at the National Institute of Standards and Technology,'' \textit{Applied Radiation and Isotopes,}} 2002, {\bf 56}, Iss. 1-2, 125-130. \textbf{Topics: D,E}

\item Unterweger M. P., Fitzgerald R., \href{https://doi.org/10.1016/j.apradiso.2013.11.017}{``Update of NIST half-life results corrected for ionization chamber source-holder instability,'' \textit{Applied Radiation and Isotopes,}} 2014, {\bf 87}, 92-94. \textbf{Topics: E,P}

\item van Heerden M. R., Cole K., van der Meulen N. P., Franzidis J. P., Buffler A., \href{https://doi.org/10.1016/j.apradiso.2020.109044}{``Extending the life of SnO\textsubscript{2}\textsuperscript{68}Ge/\textsuperscript{68}Ga generators used in the radiolabelling of ion exchange resins,'' \textit{Applied Radiation and Isotopes,}} 2020, {\bf 158}, 109044. \textbf{Topics: E,EC}

\item Walz K. F., \href{https://doi.org/10.1016/0020-708X(83)90187-4}{``Half-life measurements at the PTB,'' \textit{The International Journal of Applied Radiaiton and Isotopes,}} 1983, {\bf 34}, Iss. 8, 1191-1199. \textbf{Topics: E,P}

\bibitem{Ware} Ware M. J., Bergeson S. D., Ellsworth J. E., Groesbeck M., Hansen J. E., Pace D., Peatross J., \href{https://doi.org/10.1063/1.4926346}{``Instrument for precision long-term \textbeta-decay rate measurements'' \textit{Review of Scientific Instruments,}} 2015, {\bf 86}, 073505. \textbf{Topics: E,S,}

\item Wietfeld F. E., Greene G. L.,  \href{https://doi.org/10.1103/RevModPhys.83.1173}{``Colloquium: The neutron lifetime,'' \textit{Reviews of Modern Physics,}} 2011, {\bf 83,} Iss. 4, 1173-1192. \textbf{Topics: B,D,R}

\item Wink R., Anselmann P., Dörflinger D., Hampel W., Heusser G., Kirsten T., Mögel P., Pernicka E., Plaga R., Schlosser C.,  \href{https://doi.org/10.1016/0168-9002(93)91289-Y}{``The miniaturized proportional counter HD-2(Fe)/(Si) for the GALLEX solar neutrino experiment,'' \textit{Nuclear Instruments and Methods in Physics Research Section A: Accelerators, Spectrometers, Detectors and Associated Equipment,}} 1993, {\bf 329,} Iss. 3, 541-550. \textbf{Topics: E,S}

\item Woods M. J., Collins S. M., \href{https://doi.org/10.1016/j.apradiso.2003.11.026}{``Half-life data–a critical review of TECDOC-619 update,'' \textit{Applied Radiation and Isotopes,}} 2004, {\bf 60}, Iss. 2-4, 257-262. \textbf{Topics: P}

\item Yoo J., Ashie Y., Fukuda S., Fukuda Y., Ishihara K., Itow Y., Koshio Y., Minamino A., Miura M., Moriyama S.,Super Kamiokande Collaboration, et al., \href{https://doi.org/10.1103/PhysRevD.68.092002}{``Search for periodic modulations of the solar neutrino flux in Super-Kamiokande-I,'' \textit{Physical Review D,}} 2003, {\bf 68}, 092002. \textbf{Topics: E,P,S}

\item Zyla P. A., Particle Data Group, et al., \href{https://pdg.lbl.gov/2020/listings/rpp2020-list-neutrino-prop.pdf}{``Neutrino Properties,'' \textit{Progress of Theoretical and Experimental Physics,}} 2020, {\bf 2020}, 083C01. \textbf{Topics: E,P,S,T}
\end{enumerate}


\section*{References}

\begin{thebibliography}{99}
\bibitem{Alburger}  Alburger D. E., Harbottle G., Norton E. F., ``Half-life of \textsuperscript{32}Si,''  \textit{Earth and Planetary Science Letters,} 1986, {\bf 78}, Iss. 2-3, 168-176. 

\bibitem{Javorsek_intro} Javorsek D., Sturrock P. A., Lasenby R. N., Lasenby A. N., Buncher J. B., Fischbach E., Gruenwald J. T., Hoft A. W., Horan T. J., Jenkins, J. H., et al., ``Power spectrum analyses of nuclear decay rates,'' \textit{Astroparticle Physics,} 2010, {\bf 34}, Iss. 3, 173-178.

\bibitem{PTB}  Siegert H., Schrader H., Sch\"{o}tzig U.,``Half-life measurements of Europium radionuclides and the long-term stability of detectors,'' \textit{Applied Radiation and Isotopes,} 1998, {\bf 49}, Iss. 9-11, 1397-1401.  

\bibitem{Pomme} Pommé S., Stroh H., Paepen J., Van Ammel R., Marouli M., Altzitzoglou T., Hult M., Kossert K., Nähle O., Schrader H., et al. ``Evidence against solar influence on nuclear decay constants,'' \textit{Physics Letters B,} 2016, {\bf 761}, 281-286.

\bibitem{Davis} Davis Jr., R.  ``A Review of Measurements of the Solar Neutrino Flux and their Variation,'' \textit{Nucl. Phys. B (Proc. Suppl.)}, 1996, {\bf 48}, 284--298.

\bibitem{flare}  Jenkins, J. H., Fischbach E., ``Perturbation of nuclear decay rates during the solar flare of 2006 December 13,'' \textit{Astroparticle Phyiscs,} 2009, {\bf 31}, Iss. 6, 407-411. 

\bibitem{inspiral2} Fischbach E., Krause D. E., Pattermann M., Comment on ``Testing claims of the GW170817 binary star inspiral affecting $\beta$-decay rates,'' arXiv:2003.00092 [nucl-ex].

\bibitem{inspiral1} Fischbach E., Barnes V. E., Cinko N., Heim J., Kaplan H. B., Krause D. E., Leeman J. R., Mathews S. A., Mueterthies M. J., et al., ``Indications of an unexpected signal associated with the GW170817 binary neutron star inspiral,'' \textit{Astroparticle Physics,} 2018, {\bf 103}, 1-6.

\bibitem{Agafonova} Agafonova, N. Yu., Malgin, A.S., Fischbach, E. ``Relationship between detector signals recorded during events SN1987A and GW170817'', submitted to \textit{Astroparticle Physics}.

\bibitem{Mueterthies} Mueterthies  M. J.,  Krause D.E., Longman A.,  Barnes V.E,  Fischbach E.  ``Is there a signal for Lorentz non-invariance in existing radioactive decay data?'' In \textit{Proceedings of the Seventh meeting on CPT and Lorentz symmetry}, (Edited by V. A. Kosteleck\'{y}), World Scientific, Singapore, 2017, pp.~197--200.
\newpage
\end{thebibliography}
\end{document}